\documentclass[journal]{IEEEtran}
\usepackage{amssymb}
\usepackage{color,graphicx}
\usepackage{cite}
\usepackage[cmex10]{amsmath}
\usepackage{amsopn}
\usepackage{slashbox}
\usepackage{stfloats}
\usepackage{multirow}
\usepackage{enumerate}
\usepackage{bm}

\allowdisplaybreaks
\begin{document}
\title{{Precoding for the Sparsely Spread MC-CDMA Downlink with Discrete-Alphabet Inputs}}
\author{Min Li,~\IEEEmembership{Member,~IEEE}, Chunshan Liu,~\IEEEmembership{Member,~IEEE}, and Stephen V. Hanly,~\IEEEmembership{Senior Member,~IEEE} \thanks{Copyright (c) 2015 IEEE. Personal use of this material is permitted. However, permission to use this material for any other purposes must be obtained from the IEEE by sending a request to pubs-permissions@ieee.org. \par This work was presented in part in Proceedings of the IEEE International Conference on Communications (ICC), Sydney, Australia, June 2014. This research was supported in part by the Australian Research Council under grant DP130101760, and by the CSIRO Macquarie University Chair in Wireless Communications. This Chair has been established with funding provided by the Science and Industry Endowment Fund. \par The authors are with the Department of Engineering, Macquarie University, Macquarie Park, NSW 2113,
Australia (e-mail: \{min.li, chunshan.liu, stephen.hanly\}@mq.edu.au).}}

\markboth{IEEE Transactions on Vehicular Technology, VOL.*, NO.*, MONTH 2016}
{Shell \MakeLowercase{\textit{et al.}}: Bare Demo of IEEEtran.cls for Journals}
 \maketitle

\begin{abstract}
\par Sparse signatures have been proposed for the CDMA uplink to reduce multi-user detection complexity, but they have not yet been fully exploited for its downlink counterpart. In this work, we propose a Multi-Carrier CDMA (MC-CDMA) downlink communication, where regular sparse signatures are deployed in the frequency domain. Taking the symbol detection point of view, we formulate a problem appropriate for the downlink with discrete alphabets as inputs. The solution to the problem provides a power-efficient precoding algorithm for the base station, subject to minimum symbol error probability (SEP) requirements at the mobile stations. In the algorithm, signature sparsity is shown to be crucial for reducing precoding complexity. Numerical results confirm system-load-dependent power reduction gain from the proposed precoding over the zero-forcing precoding and the regularized zero-forcing precoding with optimized regularization parameter under the same SEP targets. For a fixed system load, it is also demonstrated that sparse MC-CDMA {with a proper choice of sparsity level} attains almost the same power efficiency {and link throughput} as that of dense MC-CDMA yet with reduced precoding complexity, thanks to the sparse signatures.
\end{abstract}


\section{Introduction}\label{sec:introduction}
\subsection{Motivations and Contributions}
\par Multi-Carrier Code Division Multiple Access (MC-CDMA) is a multi-access scheme based on the Orthogonal Frequency Division Multiplexing (OFDM) method. Since its invention, MC-CDMA has attracted broad interest, see, e.g., \cite{MCCDMA1993, hara1997overview,tulino2005spectral,fazel2008multi} and the references therein. MC-CDMA naturally integrates CDMA's flexible multiuser access with interference suppression capability and the advantages of multicarrier OFDM, including robustness against frequency-selective fading. Therefore, it has the potential to be one of the candidates to support massive {access} and provide {reliable data communication and better coverage} for future-generation wireless systems.
\par As in all CDMA systems, MC-CDMA may experience severe multi-access interference due to the loss of user orthogonality, which may occur, particularly, in frequency-selective channel environments. For such systems, optimal detection entails an exponential number of hypothesis testings about data symbols of all users and thus could be computationally demanding, time-consuming and even infeasible in a large system with conventional dense signature design. To circumvent the complexity issue, sparse signatures, whose fraction of non-zero entries is small, have been introduced and exploited for the CDMA uplink multi-user detection~\cite{yoshida2006analysis,montanari2006analysis,raymond2007sparsely,guo2008multiuser,hoshyar2008novel,razavi2012receiver}. In particular, the belief-propagation algorithm has been proposed for such a system, an algorithm that has a natural implementation using parallel computation units, and one that is fast and provably optimal for different ensembles of sparse signatures in the large system limits \cite{yoshida2006analysis,montanari2006analysis,raymond2007sparsely,guo2008multiuser}. {Inspired by the belief-propagation algorithm, references~\cite{hoshyar2008novel} and \cite{razavi2012receiver} have developed reduced-complexity soft-in soft-out (SISO) and Turbo iterative multiuser detection algorithms for the sparse CDMA uplink and a sparse-signature OFDM uplink, respectively.}

\par In this work, we formulate a different problem, appropriate for the MC-CDMA downlink counterpart, from the symbol detection point of view. The solution to the problem provides a power-efficient precoding algorithm for the base station (BS). {The precoding is implemented at the BS, allowing the mobile stations (MSs) to use simple conventional single-user matched filters and standard single-user symbol detection. This hence simplifies the implementation of the receiver, as compared to the conventional MC-CDMA downlink transceiver design, where multi-user interference is normally mitigated by a frequency-domain equalization at the receiver~\cite{hara1997overview}. Moreover, the proposed algorithm optimizes the signals transmitted on different subcarriers so that they can be constructively combined at each receiver, leading to a power-efficient precoding}. {The use of random sparse signatures was suggested for the MC-CDMA downlink in \cite{choi2004low} to allow low-complexity iterative multiuser detection at each MS. But here we show that sparsity can be exploited to reduce the complexity of the precoding, as compared to MC-CDMA with dense signatures. In addition, using sparse signatures simplifies channel measurement, since each MS only needs to estimate channels for a small number of subcarriers it occupies.}

\par The contributions of this paper are summarized as follows:
\begin{itemize}
  \item We introduce the MC-CDMA downlink communication with regular sparse signatures, where each MS has access to an equal number of subcarriers and each subcarrier has (roughly) equal load. We also consider a bipartite graph representation for the system studied and use it to facilitate algorithm design and complexity analysis.
  \item Assume that data symbols intended for MSs are drawn from discrete alphabets. We take the symbol detection point of view and introduce the minimum Symbol Error Probability (SEP) as a Quality of Service (QoS) metric for each MS. We formulate the precoding problem as a transmit power optimization problem subject to minimum SEP requirements at MSs. We translate the SEP targets into constraint regions on noiseless received signal components at MSs and characterize them via a conservative approximation to make the problem tractable. Detailed formulation procedures are provided for systems with both standard 4/16-QAM constellations and Tomlinson-Harashima replica points.
  \item We develop a precoding algorithm that accommodates parallel computation units via the dual-decomposition theory. Aided by the graph representation of the system, the complexity of the algorithm is characterized in terms of the number of message passings between computation units and the number of additions and multiplications for precoding calculation. Signature sparsity is shown to play a vital role in reducing precoding complexity.
  \item We demonstrate that the proposed optimized precoding generally outperforms the conventional zero-forcing (ZF) {and the optimized Regularized ZF (RZF) precodings} in terms of power efficiency under the same SEP requirements. The exact gain depends on the system load and it is very significant for a fully loaded system. We also demonstrate that, for a fixed system load, sparse MC-CDMA, e.g., with {a proper relatively small number} of subcarriers allocated to each MS, attains almost the same power efficiency {and link throughput} as dense MC-CDMA under our proposed precoding scheme. This important observation, in conjunction with the fact that sparsity reduces precoding complexity, promotes the practicality of sparse MC-CDMA.
\end{itemize}

\subsection{Other Related Work}
\par Precoding is a relatively mature concept in multi-antenna communication systems, enabling multiuser multiplexing in the spatial domain, see, e.g., \cite{peel2005vector,hochwald2005vector,wiesel2006linear,sanguinetti2007non,muller2008vector,shenouda2009nonlinear,kang2009tomlinson,garcia2014power,bjornson2014optimal}. It has to balance between the two conflicting interests of maximizing the useful signal power at the intended user and minimizing interference leakage towards non-intended users. The same concept can be applied to other systems such as direct-sequence CDMA or MC-CDMA, where multiplexing takes place in the time- or frequency-domain, and multi-access interference, if it arises, has to be dealt with \cite{silva2003pre,salzer2003,cosovic2005non,morelli2007unified,masouros2010soft}.
\par Existing precoding techniques can be divided into two categories, linear precoding and non-linear precoding, where both {require channel state information while the latter requires additional symbol-based processing}. Within the first category, matched filtering, ZF and RZF \cite{peel2005vector} are three commonly known precodings that maintain different levels of balance between the two conflicting goals. Other power-aware linear precodings of general form have also been proposed in the literature subject to different QoS metrics, such as signal-to-noise-plus-interference ratio~\cite{bjornson2014optimal}.
\par Compared with linear precoding, non-linear precoding may offer higher power efficiency and transmission rate, but the gain comes at the cost of incorporating more sophisticated signal processing \cite{hochwald2005vector,sanguinetti2007non,muller2008vector,shenouda2009nonlinear,cosovic2005non,morelli2007unified,masouros2010soft}. In the multi-antenna broadcasting setup, capacity-achieving non-linear Dirty-Paper Coding (DPC) entails a successive pre-cancelation of known intra-user interference at the BS. The encoding of data relies on codewords of infinite length and involves a high-dimensional sphere-search algorithm, which renders DPC unattractive in practical systems. Tomlinson-Harashima precoding (THP) is a simplified version of DPC, where the codebook is comprised of periodic extension of standard constellations (replica points) in the two-dimensional space and a transmit modulo-operation is introduced in the interference pre-cancelation process in order to reduce transmit power. Built on ZF or RZF, reference~\cite{hochwald2005vector} generalizes the single-user-based symbol extension idea of THP and introduces a joint perturbation of user symbol vector to further reduce transmit power.
\par The optimized precoding proposed in this work belongs to the second category. As in existing works, power consumption is one of the primary concerns in our optimized precoding. However, the optimization criterion, minimum SEP constraint, has not been considered before, except our own works \cite{isit2013,ita2013,li2014distributed} in the MIMO (or distributed MIMO) setup. This criterion appears to be natural when we consider a system with discrete alphabets as inputs. In addition, in our formulation, we fix the information-bearing alphabets at the BS but allow a certain relaxation of received signals at MSs through precoding, as long as they reside in detection-favorable regions and the SEP targets are met. This is distinguished from related works \cite{muller2008vector,kang2009tomlinson,garcia2014power,masouros2010soft}, where non-linear relaxation of input alphabets has been adopted. In \cite{muller2008vector,kang2009tomlinson}, the relaxation is required to maintain the minimum signalling distance, while in \cite{masouros2010soft,garcia2014power}, the relaxation is to ensure the corresponding symbol-energy-to-noise ratio is above a certain threshold \cite{masouros2010soft,garcia2014power}. However, in all these works, no explicit SEP targets are imposed at MSs.

\par {\it Notation}: Boldface uppercase and lowercase letters denote matrices and vectors, respectively, e.g., $\mathbf{A}$ is a matrix and $\mathbf{a}$ is a vector; ${\bf I}_N$ is an $N \times N$ identity matrix; for integers $i \le j$, $[i:j] = \{i,i+1,\dots,j\}$, is the discrete interval between $i$ and $j$, and $\mathbf{a}_{[i:j]} = \{a_i,\dots,a_{j}\}$, is the collection of $[i:j]$th components of vector $\mathbf{a}$; $(\cdot)^T$ denotes the matrix transpose, while $(\cdot)^\dag$ denotes the conjugate transpose; notation ${\mathsf E}[X]$ denotes the expectation operation on random variable $X$, and $\left \lfloor {x} \right \rfloor$ denotes a floor function of real number $x$; $\Re\{\cdot\}$ and $\Im\{ \cdot\}$ denote the real and imaginary part of a complex number, respectively; finally, $\|\mathbf{A}\|_p$ is the standard $l_p$ norm of $\mathbf{A}$.
%\vspace{-0.2em}

\section{System Model}\label{sec:system:model}

\subsection{Signaling Model}
\par We consider a downlink communication, where a single-antenna BS is simultaneously serving $K$ single-antenna MSs via MC-CDMA. Specifically, data symbols intended for MSs are all drawn from discrete alphabet sets, e.g., $M$-QAM constellations, common in practical deployments. The downlink communication takes place over a set of $N$ orthogonal subcarriers where we assume $N \ge K$ and thus the load $\alpha={K}/{N} \in \left(0,1\right]$. In the conventional MC-CDMA, the data symbol intended for a MS is transmitted over all parallel subcarriers where each is encoded with a binary phase-offset~\cite{MCCDMA1993}. Here, however, information associated with each MS is assumed to be spread onto only a small subset of the available subcarriers, which leads to the sparse MC-CDMA as originally studied by \cite{yoshida2006analysis,montanari2006analysis,raymond2007sparsely,guo2008multiuser} for the uplink.

\par Let ${\bf s}_k = \frac{1}{\sqrt{L_k}}\left[{\tilde s}_{k,1},{\tilde s}_{k,2},\ldots,{\tilde s}_{k,N}\right]^T$ be the signature for MS $k$, where normalization factor $L_k$ corresponds to the total number of subcarriers allocated to MS $k$. In the signature, components ${\tilde s}_{k,n}$ are i.i.d. drawn from a distribution $P_S$ with zero-mean and unit-variance if MS $k$ has access to subcarrier $n$, and ${\tilde s}_{k,n}= 0$ otherwise. The collection of all signatures corresponds to a sparse signature matrix ${\bf S} = \left[ {{\bf s}_1, {\bf s}_2, \ldots, {\bf s}_K} \right]$, which is perfectly known at the BS.
\par
\begin{figure}[t]
\centering
\includegraphics[width=0.5\textwidth]{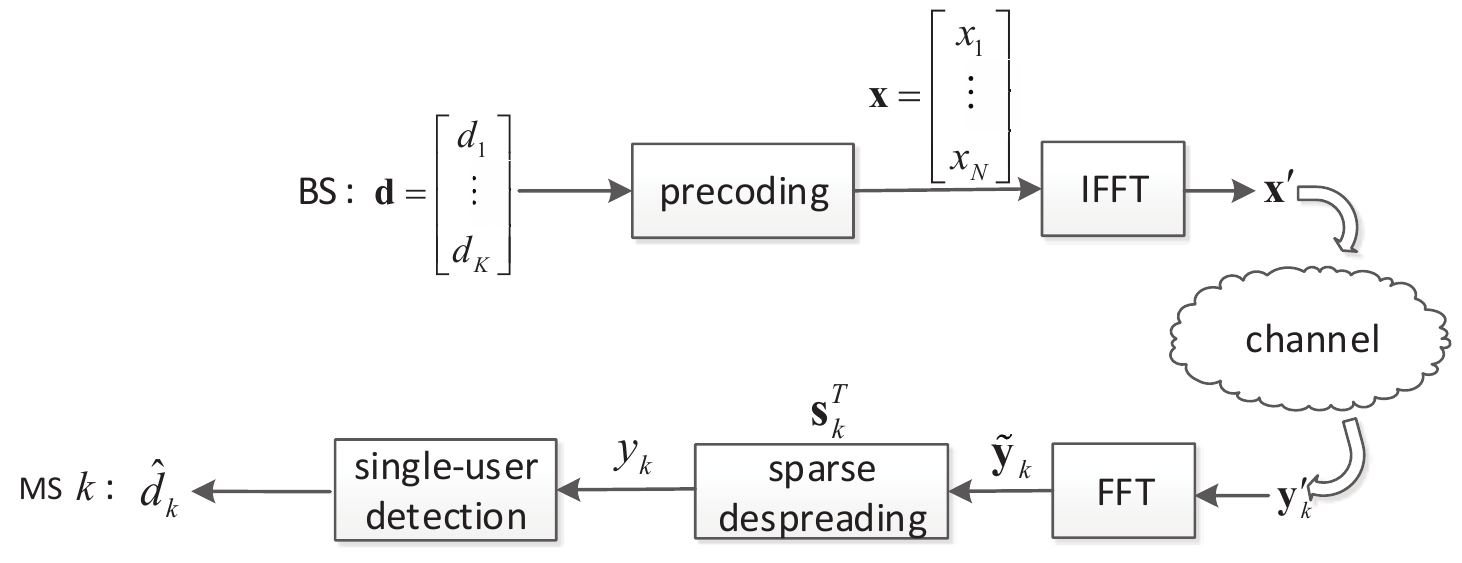}
\caption{The transceiver architecture for the sparse MC-CDMA downlink.}\label{fig:transceiver}
\end{figure}

 The transceiver architecture for the downlink transmission is depicted in Fig.~\ref{fig:transceiver} and is elaborated as follows. Let ${\bf d} = \left[d_1,\ldots,d_K\right]^T$ be the transmitted data symbol vector, where component $d_k$ denotes the symbol intended for MS $k$ and is drawn from a discrete finite-alphabet set ${\cal D}_k$. The transmission takes place by first forming appropriate frequency-domain signals and then converting them into time-domain signals by the inverse fast Fourier transform (IFFT) at the BS. Specifically, symbol vector ${\bf d}$ is passed through a precoder and mapped to coded vector ${\bf x} \in {\mathbb C}^N$. An IFFT is then applied over the coded vector in order to generate time-domain signal vector ${\bf {x}}{'} \in {\mathbb C}^N$ that is subsequently transmitted over the wireless channel.

\par Upon observing channel output, each MS $k$ first performs an FFT and produces the frequency domain signal ${\bf \tilde y}_k$ as
\begin{align}
{\bf \tilde y}_{k} = {\bf \tilde h}_{k} \circ {\bf x} + {\bf \tilde z}_{k},
\end{align}
where notation ``$\circ$'' denotes the Hadamard product; vector ${\bf \tilde h}_{k} = [{\tilde h}_{k,1},{\tilde h}_{k,2},\ldots,{\tilde h}_{k,N}]^T$ is the collection of frequency-domain channel gains from BS to MS $k$; vector ${\bf \tilde z}_{k}$ is a circularly symmetric complex Gaussian noise with ${\mathsf E}[{\bf \tilde z}_{k}{\bf  \tilde z}_{k}^{\dag}] = N_0 {\bf I}_N$. Despreading is then performed at MS $k$ based on its own signature ${\bf s}_k$, followed by a simple single-user detection. The corresponding output signal $y_k$ after despreading is given by
\begin{align}
y_k  = \sum\limits_{n = 1}^N {\underbrace {\left( {s_{k,n} \tilde h_{k,n} } \right)}_{h_{k,n} }} x_n  + \underbrace {{\bf{s}}_k^T {\bf{\tilde z}}_k }_{z_k } = \sum\limits_{n = 1}^N {h_{k,n} } x_n  + z_k,
\end{align}
where $s_{k,n}={\tilde s}_{k,n}/\sqrt{L_k}$ is the $n$th component of ${\bf s}_k$, and each equivalent channel noise ${z}_k$ is circularly symmetric complex Gaussian with zero mean and variance $N_0$. Collecting all outputs at MSs, we obtain the equivalent system input-output relationship as
{\footnotesize{ \begin{align}
\underbrace {\left[ {\begin{array}{*{20}c}
   {y_1 }  \\
   {y_2 }  \\
    \vdots   \\
   {y_K }  \\
\end{array}} \right]}_{{\bf{y}}} = \underbrace {\left[ {\begin{array}{*{20}c}
   {h_{1,1} } &  \cdots  &  \cdots  & {h_{1,N} }  \\
   {h_{2,1} } &  \cdots  &  \cdots  & {h_{2,N} }  \\
    \vdots  &  \vdots  &  \vdots  &  \vdots   \\
   {h_{K,1} } &  \cdots  &  \cdots  & {h_{K,N} }  \\
\end{array}} \right]}_{{\bf{H}}}\underbrace {\left[ {\begin{array}{*{20}c}
   {x_1 }  \\
   {x_2 }  \\
    \vdots   \\
   {x_N }  \\
\end{array}} \right]}_{\bf{x}} + \underbrace {\left[ {\begin{array}{*{20}c}
   {z_1 }  \\
   {z_2 }  \\
    \vdots   \\
   {z_K }  \\
\end{array}} \right]}_{{\bf{z}}}. \label{equ:system:H}
\end{align}}}
It is straightforward to observe that in matrix ${\bf H}$, $h_{k,n} =0$ as long as $s_{k,n} =0$, and thus row ${\bf h}_k$ maintains the same level of sparsity as the corresponding signature ${\bf s}_k$. This also means the BS only needs to know the small number of $h_{k,n}$s for which $s_{k,n} \ne 0$ for the purpose of precoding.

\begin{figure}[t]
\centering
\includegraphics[width=0.25\textwidth]{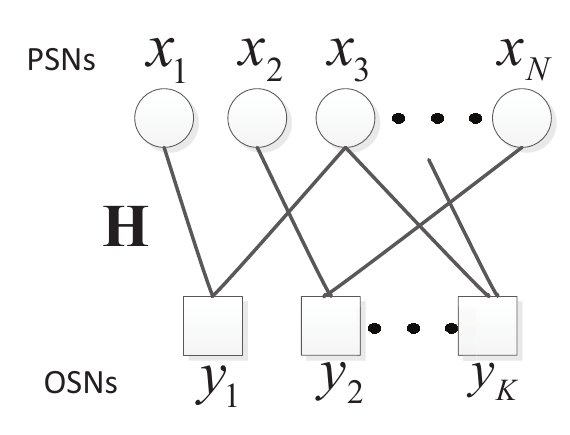}
\caption{Bipartite graph representation of the sparse MC-CDMA downlink.}\label{fig:graph:representation}
\end{figure}

\subsection{Graph Representation}
\par Given the matrix ${\bf H}$ from~\eqref{equ:system:H}, we can alternatively construct a bipartite graph representation of the sparse MC-CDMA system. Assume that each symbol $x_n$ in the graph is represented by a precoded symbol node (PSN), and each output $y_k$ is represented by a output symbol node (OSN). PSNs will be drawn as circles and OSNs will be drawn as squares in the graph. PSN $x_n$ is connected with OSN $y_k$ only if $s_{k,n} \ne 0$ and $h_{k,n}$ is the weight associated with the edge. Fig.~\ref{fig:graph:representation} depicts an instance of the graph ${\cal G}$ for $L_k=2$, where each OSN is connected with two PSNs. We use ${\cal I}(x_n)$ to denote the collection of OSNs connected to $x_n$ and define the node degree of $x_n$ as the cardinality $\left|{\cal I}(x_n)\right|$. Similarly, we use ${\cal I}(y_k)$ to denote the collection of PSNs connected to $y_k$ and define the node degree of $y_k$ as the cardinality $\left|{\cal I}(y_k)\right|$. This graph representation introduced will facilitate the description of the precoding algorithm and the corresponding complexity analysis in Section~\ref{sec:precoding:algorithm}.

\subsection{Sparse Signature Ensemble} \label{subsec:signature}
\par In the signature matrix, we assume that the non-zero elements $\{{\tilde s}_{k,n}\}$ are i.i.d. drawn from a uniform distribution on $\{+1, -1\}$. {It is observed that generating non-zero elements according to other distributions, e.g., Gaussian distribution, has little impact on the averaged system performance. Hence, we stick to the binary uniform distribution, which leads to a binary phase-offset for the symbol transmitted as in \cite{montanari2006analysis,raymond2007sparsely}}.
\par Depending on the number of subcarriers allocated across MSs and the load per subcarrier, we have three common signature ensembles suggested for the uplink \cite{raymond2007sparsely}: {\it i}) irregular ensemble, where Poisson-distributed number of subcarriers are allocated across MSs and the load per subcarrier is also Poisson-distributed; {\it ii}) semi-regular ensemble, where each MS is allocated a fixed positive integer number of subcarriers and the load per subcarrier is Poisson-distributed; and {\it iii}) regular ensemble, where the number of subcarriers allocated for each MS and the loading per subcarrier both take fixed positive integer values. In particular, \cite{raymond2007sparsely} advocates the regular ensemble as it amongst others prevents the systematic inefficiency due to leaving some subcarriers unoccupied by any of MSs.
\par In this work, we follow \cite{raymond2007sparsely} and deploy the regular-type ensemble {to ensure that the system enjoys full utilization of resources and provides user fairness}. Specifically, when the system is fully-loaded ($\alpha=1$), a perfectly regular signature matrix is randomly generated in the sense that each MS is allocated $L$ subcarriers, and each subcarrier is accessed by exactly $L$ MSs. When the system is under-loaded ($\alpha < 1$), a nearly regular signature matrix is randomly generated such that each MS is allocated $L$ subcarriers, and each subcarrier is of almost equal load, namely, accessed by either $\left \lfloor {\alpha L} \right \rfloor$ or $ (\left \lfloor {\alpha L} \right \rfloor+1)$ MSs.

\section{Optimized Precoding with SEP Targets} \label{sec:problem:formulation}
\par For notational convenience, we define $\overline{y}_k = \mathbf{h}^T_k\mathbf{x} $ as the noiseless received component at MS $k$ ($\mathbf{h}^T_k$ is the $k$th row of matrix ${\bf H}$) with real part $\overline{y}^{(r)}_k = \Re\{\overline{y}_k\}$ and imaginary part $\overline{y}^{(i)}_k = \Im\{\overline{y}_k\}$; similarly, denote the real and imaginary parts of data symbol and noise as $d^{(r)}_k = \Re\{d_k\}$ and $d^{(i)}_k = \Im\{d_k\}$, and $z^{(r)}_k = \Re\{z_k\}$ and $z^{(i)}_k = \Im\{z_k\}$, respectively. In addition, define $\sigma^2 = N_0/{2}$ as the fixed noise variance per signal dimension.
\par Assuming that all data symbols transmitted are selected from discrete alphabets, we take a symbol detection point of view and impose minimum Symbol Error Probabilities (SEPs) as the user QoS constraints. Specifically, let ${\cal A}(d_k)$ denote the decision region associated with data symbol $d_k$ intended for MS $k$ and $Pe_k$ denote the SEP target. Detection error happens when the output signal $y_k$ lies outside decision region ${\cal A}(d_k)$. According to the SEP requirement, the probability of error events should be no greater than the target, i.e.,
\begin{align} \label{Eq:SEP}
\Pr\left(y_k=\left(\bar y_k + z_k\right) \not\in \mathcal{A}(d_k)\right) \le Pe_k.
\end{align}
\par A question we then ask is: How do we design a precoder that efficiently maps a symbol vector ${\bf d}$ into a precoded vector ${\bf x}$ such that the SEP requirements at MSs can be met?

\par Following the conventional zero-forcing (ZF) approach, one could form ${\bf x}$ according to
\begin{align}
{\bf x} = {\bf H}^\dag\left({\bf H}{\bf H}^\dag\right)^{-1}{\bf d}, \label{equ:conventional:zf}
\end{align}
which inverts the channel matrix and forces noiseless component $\bar y_k$ to sit exactly at the constellation point $d_k$. In order to meet a given SEP target, $Pe_k$, data symbol $d_k$ has to be chosen from a discrete alphabet set whose minimum distance between any two neighboring points (denoted by ${\sf d}_{\text {min}}$) is above a certain threshold.
\par For instance, consider a system with $M$-QAM modulation whose standard constellation is represented by
\begin{align}
{\cal D}_{\text S} = { \left\{a_{\rm R} + {j}a_{\rm I}\left|a_{\rm R},a_{\rm I}\in \{\pm 1, \pm 3, \dots,\pm ({\sqrt M}-1)\}\right.\right\}}
\end{align}
with ${\sf d}_{\text {min}} = 2$. We need to scale the constellation points (increasing the minimum distance ${\sf d}_{\text {min}}$, but also the transmit power) in order to meet the SEP target, $Pe_k$. Considering the 4-QAM constellation (see Fig.~\ref{fig:4:QAM}), the scaling factor, $\beta_k$, to use for MS $k$, must satisfy
\begin{align}
{\frac{1}{\sqrt{2\pi}\sigma}\int_{-\beta_k}^{+\infty}e^{-\frac{\left( z^{(r)}_k \right)^2}{2\sigma^2}} dz^{(r)}_k } &\times {\frac{1}{\sqrt{2\pi}\sigma}\int_{-\beta_k}^{+\infty}e^{-\frac{\left( z^{(i)}_k \right)^2}{2\sigma^2}} dz^{(i)}_k} \nonumber\\
&\quad \quad \quad \quad \quad \quad \footnotesize{\geq 1-Pe_k,}
\end{align}
as implied by \eqref{Eq:SEP}. Thus the minimum scaling factor under the conventional ZF approach for the 4-QAM system is given by
\begin{align}
\beta_k^- = -{\sigma}{Q}^{-1}\left(\sqrt{1-Pe_k}\right), \label{equ:beta:minimum}
\end{align}
where $Q^{-1}(.)$ denotes the inverse of the standard $Q$-function \cite{proakis2007}. When $M \ge 16$ (see Fig.~\ref{fig:16:QAM} for 16-QAM), the standard constellation ${\cal D}_{\text S}$ should be scaled so that
\begin{align}
{\frac{1}{\sqrt{2\pi}\sigma}\int_{-\beta_k}^{+\beta_k}e^{-\frac{\left( z^{(r)}_k \right)^2}{2\sigma^2}} dz^{(r)}_k } &\times {\frac{1}{\sqrt{2\pi}\sigma}\int_{-\beta_k}^{+\beta_k}e^{-\frac{\left( z^{(i)}_k \right)^2}{2\sigma^2}} dz^{(i)}_k} \nonumber \\
&\quad \quad \quad \quad \quad \quad\footnotesize{\geq 1-Pe_k,}
\end{align}
as implied by \eqref{Eq:SEP}, considering the dominant scenario in which one of the inner most points is transmitted. Thus the minimum scaling factor under the conventional ZF approach is given by
 \begin{align}
\beta_k^- = {\sigma}{Q}^{-1}\left(0.5-0.5\sqrt{1-Pe_k}\right). \label{equ:beta:minimum:16QAM}
\end{align}
\par In general, however, we do not have to zero-force ${\bar y}_k$, and in fact, it is sufficient to ensure ${\bar y}_k$ falls into a region that favours correct symbol detection. This relaxation introduces room to optimize the choice of ${\bf x}$, leading to the following transmit power minimization problem:
\begin{align}
{\bf \cal P}:\left\{ {\begin{array}{*{20}l}
   {\min \limits_{\mathbf{x} \in {\mathbb C}^{N \times 1}} P\left({\bf x}\right) = \mathbf{x}^\dag\mathbf{x} }  \\
   {\text{subject to\:\:}\Pr\left(\left({\bar y}_k + z_k\right) \not\in \mathcal{A}(d_k)\right) \le Pe_k,}\\
   {{\text{for the transmitted data set\:}} \{d_k \in {\cal D}_k,\;k = 1, \ldots, K\}}.  \\
\end{array}} \right. \label{Problem:general}
\end{align}

\section{Sparse MC-CDMA with Standard $M$-QAM Constellations}\label{sec:standard:constellation}
\par In this section, we show how to translate the set of SEP targets in~\eqref{Problem:general} into constraints on noiseless output components at MSs. In particular, we begin with the 4-QAM signaling case and then generalize to the 16-QAM signaling case. A similar approach can be applied to systems with {higher-order QAM} constellations.

\subsection{Translate SEP Targets to Constraints on Noiseless Received Signal Components}

\subsubsection{4-QAM} \label{subsubsec:4QAM}
 \par
\begin{figure}[t]
\centering
\includegraphics[width=0.32\textwidth]{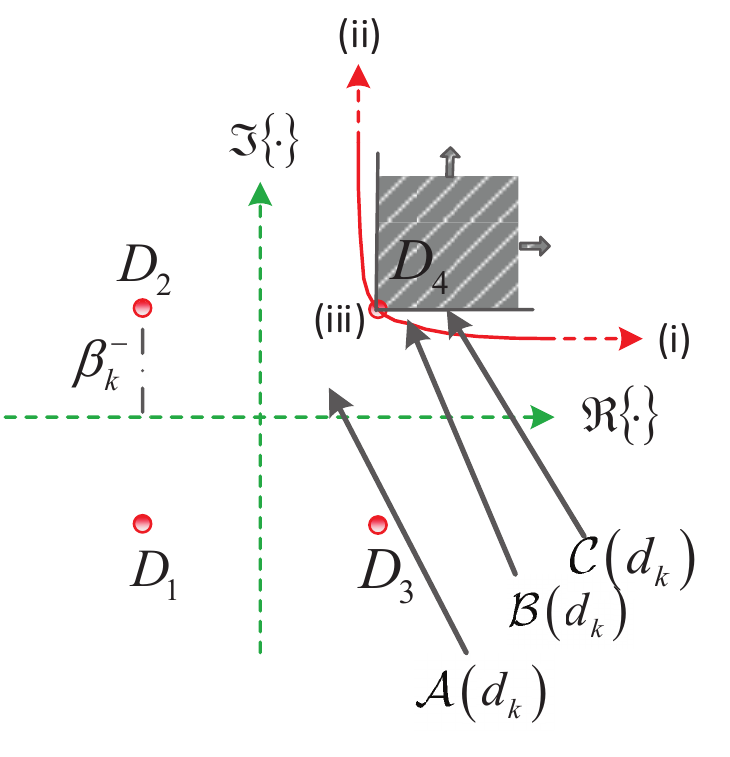}
\caption{4-QAM: $\beta_k^-$ is the minimum scaling factor under ZF; $\{D_1, D_2, D_3, D_4\}$ are constellation points; taking $d_k = D_4$ as an example, ${\cal A}(d_k)$ corresponds to the decision region, ${\cal B}(d_k)$ corresponds to the precise constraint region on noiseless output $\overline{y}_m$, while ${\cal C}(d_k)$ represents the constraint region with conservative approximation.}\label{fig:4:QAM}
\end{figure}

 Assume that each $d_k$ is drawn from the 4-QAM constellation set ${\cal{D}}_{k}=\{D_m:m=1,\dots,4\}$ as shown in Fig.~\ref{fig:4:QAM}, where the green dashed lines partition the complex plane into four symmetric decision regions each occupying an open quarter plane. Any received signals falling outside the correct region lead to detection error. Thus, the SEP requirement \eqref{Eq:SEP} becomes
\begin{align}\label{Cons:QAM}
&\underbrace{\frac{1}{\sqrt{2\pi}\sigma}\int_{I^{(r)}_{-}(d_k)}^{I^{(r)}_{+}(d_k)}e^{-\frac{\left( z^{(r)}_k \right)^2}{2\sigma^2}} dz^{(r)}_k }_{\mathcal{O}^{(r)}} \nonumber  \\
&\quad \quad \times \underbrace{\frac{1}{\sqrt{2\pi}\sigma}\int_{I^{(i)}_{-}(d_k)}^{I^{(i)}_{+}(d_k)}e^{-\frac{\left( z^{(i)}_k \right)^2}{2\sigma^2}} dz^{(i)}_k}_{\mathcal{O}^{(i)}} \geq 1-Pe_k,
\end{align}
where the tuple $(I^{(r)}_{-}(d_k),I^{(r)}_{+}(d_k),I^{(i)}_{-}(d_k),I^{(i)}_{+}(d_k))$ depends on the data transmitted and equals:
\begin{enumerate}[(i)]
\item ${(-\infty, -\overline{y}^{(r)}_k,-\infty,-\overline{y}^{(i)}_k)}$ for $D_1$;
\item $(-\infty,-\overline{y}^{(r)}_k,-\overline{y}^{(i)}_k, +\infty)$ for $D_2$;
\item $(-\overline{y}^{(r)}_k, +\infty,-\infty,-\overline{y}^{(i)}_k)$ for $D_3$;
\item $(-\overline{y}^{(r)}_k,+\infty,-\overline{y}^{(i)}_k,+\infty)$ for $D_4$.
\end{enumerate}

\par Given symbol $d_k \in {\cal D}_k$ and a target $Pe_k$, one can determine the precise constraint region ${\cal B}\left(d_k\right)$ on noiseless output $\overline{y}_k $ at MS $k$ from inequality~\eqref{Cons:QAM}. In particular, the boundary of the region is determined by equality in \eqref{Cons:QAM}. For example, when $d_k = D_{4}$, three points on the boundary of ${\cal B}\left(d_k\right)$ are identified by considering combinations of $\left(\mathcal{O}^{(r)},\mathcal{O}^{(i)}\right)$:
\begin{enumerate}[(i)]
\item {$\left(1,1-Pe_k\right)$: $\overline{y}^{(r)}_k = +\infty$, $\overline{y}^{(i)}_k = - {\sigma}{Q}^{-1}(1-Pe_k)$;}
\item{$\left(1-Pe_k, 1\right)$: $\overline{y}^{(r)}_k = - {\sigma}{Q}^{-1}(1-Pe_k)$, $\overline{y}^{(i)}_k = +\infty$;}
\item{$\left(\sqrt{1-Pe_k},\sqrt{1-Pe_k}\right)$}: \\
$~~~~~~~~~~~~~~~~~~~~\overline{y}^{(r)}_k = \overline{y}^{(i)}_k = - {\sigma}{Q}^{-1}\left(\sqrt{1-Pe_k}\right).$
\end{enumerate}
In principle, the curved-shape boundary can be determined by traversing all possible combinations of $\mathcal{O}^{(r)}$ and $\mathcal{O}^{(i)}$. The constraint region ${\cal B}(d_k)$ then includes all the points on and within the boundary (see the red curve in Fig.~\ref{fig:4:QAM}). Note that point $({\text {iii}})$ is exactly a scaled constellation point. Also note that this region generally implies non-linear constraints on input signal ${\bf x}$, which makes the optimization problem less tractable. Alternatively, we can find a polytype contained in ${\cal B}(d_k)$, i.e., we conservatively approximate the region using the area bounded by line segments between a finite number of points on or within boundary. A simple approximation for $d_k= D_4$ is given by: $\overline{y}^{(r)}_k\geq - {\sigma}{Q}^{-1}\left(\sqrt{1-Pe_k}\right)$ and $\overline{y}^{(i)}_k\geq - {\sigma}{Q}^{-1}\left(\sqrt{1-Pe_k}\right)$, which lead to linear constraints on input signal $\mathbf{x}$. This region with conservative approximation is denoted as ${\cal C}(d_k=D_4)$, see the open shadow area in Fig.~\ref{fig:4:QAM}.
\par By the same approach, relaxed constraint regions associated with the other constellation points can be derived and characterized as:
 \begin{enumerate}[(i)]
 \item ${\cal C}(d_k= D_1) = \left\{(\overline{y}^{(r)}_k, \overline{y}^{(i)}_k): \overline{y}^{(r)}_k \le -I, \overline{y}^{(i)}_k \le -I \right\}$;
 \item ${\cal C}(d_k=D_2) =\left\{(\overline{y}^{(r)}_k, \overline{y}^{(i)}_k): \overline{y}^{(r)}_k \le -I, -\overline{y}^{(i)}_k \le -I\right\}$;
 \item ${\cal C}(d_k=D_3) = \left\{(\overline{y}^{(r)}_k, \overline{y}^{(i)}_k): -\overline{y}^{(r)}_k \le -I, \overline{y}^{(i)}_k \le -I\right\}$,
 \end{enumerate}
 with definition $I = - {\sigma}{Q}^{-1}\left(\sqrt{1-Pe_k}\right)$. Note that the exact areas of these regions only depend on the SEP targets.

\subsubsection{16-QAM}\label{subsubsec:16QAM}
\par
\begin{figure}[t]
\centering
\includegraphics[width=0.35\textwidth]{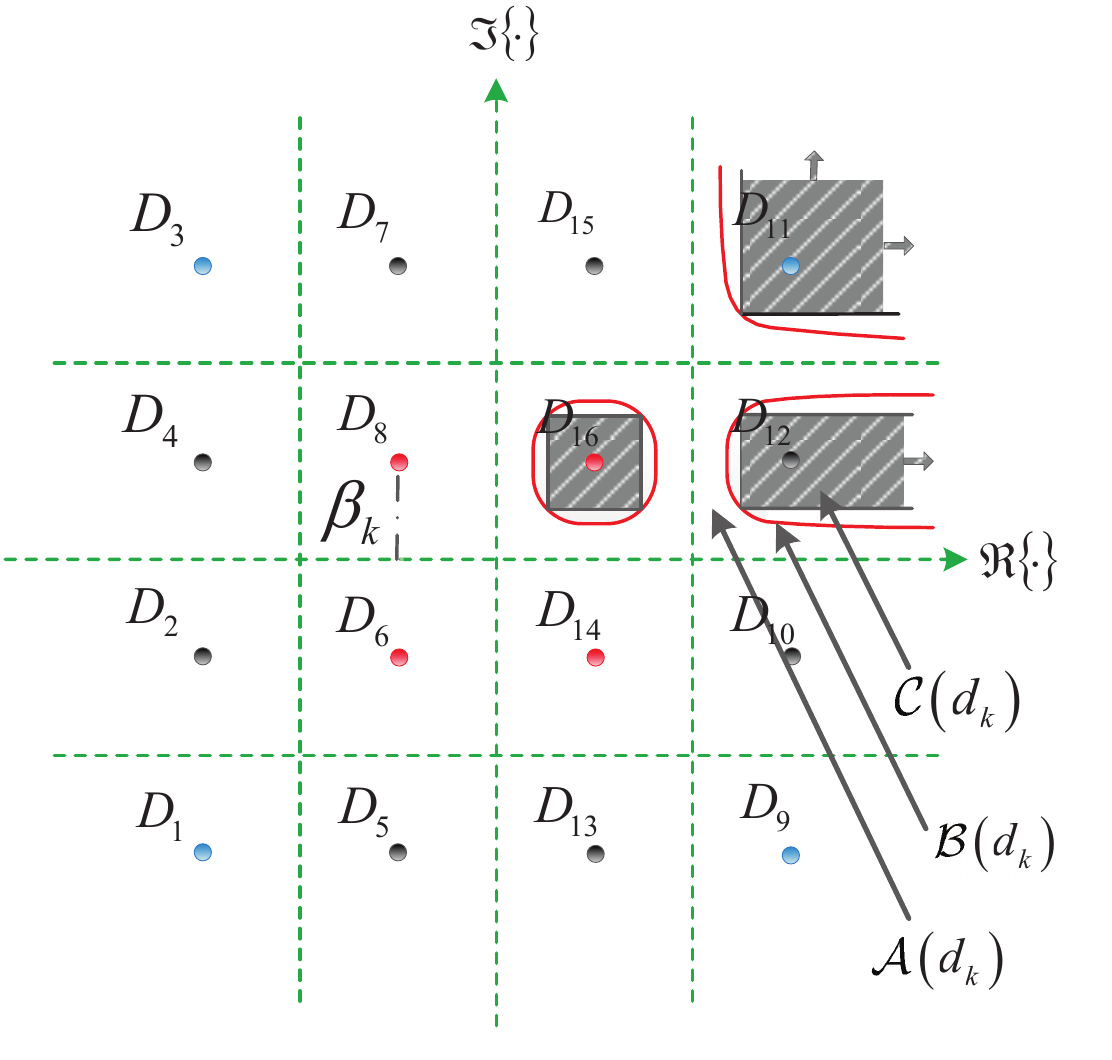}
\caption{16-QAM: $\beta_k$ is a scaling factor; $\{D_1,\ldots, D_{16}\}$ are constellation points; taking $d_k = D_{12}$ as an example, ${\cal A}(d_k)$ corresponds to the decision region, ${\cal B}(d_k)$ corresponds to the precise constraint region on noiseless output $\overline{y}_k$, while ${\cal C}(d_k)$ represents the constraint region with conservative approximation.}\label{fig:16:QAM}
\end{figure}

 We now turn to the case where $d_k$ is drawn from the 16-QAM constellation set ${\cal{D}}_{k}=\left\{D_m: m=1,\dots,16\right\}$ as shown in Fig.~\ref{fig:16:QAM}. The constraint region on $\overline{y}_k$ can be characterized based on procedures similar to those for the 4-QAM case. But calculation requires some care, since the decision regions ${\cal A}(d_k)$ for inner points and outer points are of different shapes (see the regions partitioned by the green dashed lines in Fig.~\ref{fig:16:QAM}) and the exact areas of these regions depend on the scaling factor $\beta_k$. For the sake of conciseness, the derivation of the constraint region ${\cal B}(d_k)$ is deferred to Appendix~\ref{appendix:16QAM}. Taking symbols $\left\{D_{11}, D_{12},D_{16}\right\}$ as examples, we plot the resulting regions ${\cal B}\left(d_k\right)$ in Fig.~\ref{fig:16:QAM}. Again, one can approximate these regions with a polytype ${\cal C}(d_k)$ for each, leading to linear constraints on input signal ${\bf x}$.

\begin{figure}[t]
\centering
\includegraphics[width=0.4\textwidth]{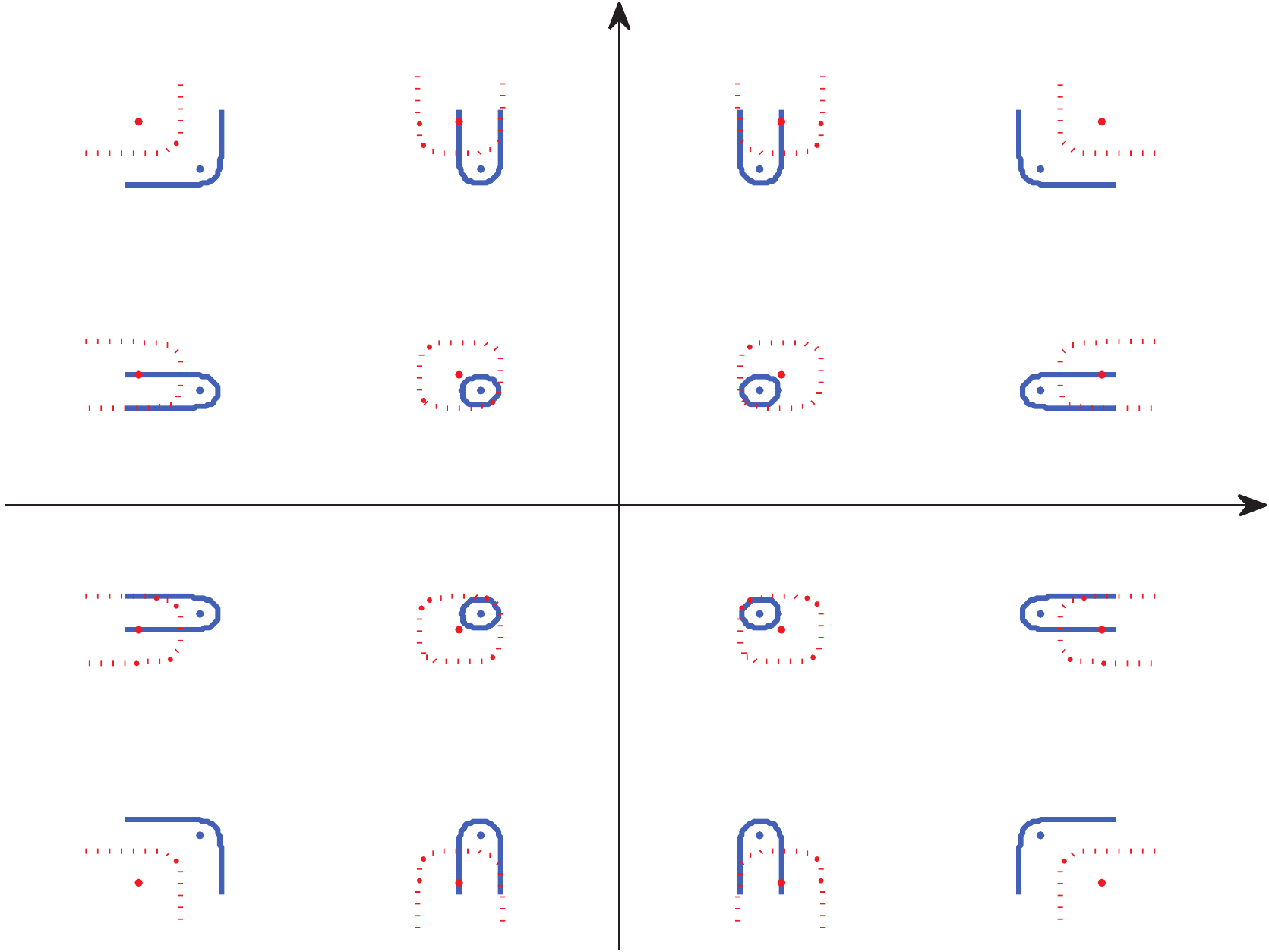}
\caption{Constraint regions ${\cal B}(d_k)$ for the 16-QAM constellation under different scaling factors with $Pe_k = 10^{-3}$ fixed: ({\it i}) blue curves: $\beta_k = 1.05\beta_k^{-}$; ({\it ii}) red curves: $\beta_k = 1.20 \beta_k^{-}$.}\label{fig:constraint_overlap}
\end{figure}

\par In general, the exact areas of ${\cal C}(d_k)$ depend on the scaling factor $\beta_k$. It is clear that, with minimum $\beta_k^-$ as defined by \eqref{equ:beta:minimum:16QAM}, constraint region ${\cal C}(d_k)$ corresponds to a strict equality constraint on $\overline{y}_k$, i.e., $\overline{y}_k = D_m$, for the center points $m\in\{6,8,14,16\}$; for the side points, e.g., $D_{12}$, ${\cal C}(d_k)$ shrinks to a line: $-\overline{y}^{(r)}_k \le -2\beta_k^{-} - I$ and $\overline{y}^{(i)}_k = \beta_k^{-}$, which implies doing a zero-forcing for the imaginary part while having a relaxed constraint on
the real part; and for the corner points, e.g., $D_{11}$, ${\cal C}(d_k)$ becomes: $-\overline{y}^{(r)}_k \le -2\beta_k^- - I, -\overline{y}^{(i)}_k \le -2\beta_k^- - I$, where we recall that $ I= - {\sigma}{Q}^{-1}\left(\sqrt{1-Pe_k}\right)$.
\par To meet a given SEP target $Pe_k$ for MS $k$, one could certainly adopt a scaling factor larger than the minimum $\beta_k^-$ for the transmission. But such a choice may affect the power efficiency of the system. Fig.~\ref{fig:constraint_overlap} plots two instances of constraint regions ${\cal B}(d_k)$ for $Pe_k=10^{-3}$ when the constellation is scaled up with $\beta_k = 1.05\beta_k^{-}$ (blue curves) and  $\beta_k = 1.20\beta_k^{-}$ (red curves). It can be seen that with a larger $\beta_k$ above $\beta_k^{-}$, when a center constellation point is transmitted, the constraint region is relaxed from a single point to a circle-type region centered on the symbol. Potential benefits can be accrued from the resulting enlarged feasible region. However, when the corner points are transmitted, the corresponding constraint regions always shrink as the constellation is scaled up. In this case, power efficiency loss may be induced because of the reduced feasible optimization space. When one of the side points is transmitted, it is unclear how the performance reacts, as the constraint region with larger $\beta_k$ partially overlaps with that for a smaller $\beta_k$. Nevertheless our experiments have indicated that when the transmitted data symbols are randomly and uniformly generated, scaling up the constellation with $\beta_k > \beta_k^{-}$ brings little further power saving. Hence we will use the minimum scaling, $\beta_k^{-}$, for the standard 16-QAM in what follows.

\subsection{Problem Reformulation with Conservative Approximation}
\par With the conservative approximation, the SEP constraints in~\eqref{Problem:general} can be translated into a set of linear inequality/equality constraints on vector ${\bf x}$. We now present the resulting optimization problem.
\par For ease of exposition, we stack the real and the imaginary parts of each $x_n$ into a real vector $\tilde{\mathbf{x}}$, i.e., $\tilde{\mathbf{x}} =\left [\Re{\{x_1\}}, \Im{\{x_1\}},\ldots, \Re{\{x_N\}}, \Im{\{x_N\}}\right]^T \in {\mathbb R}^{2N \times 1}$. For the $4$-QAM signaling, the real and imaginary parts of coded signal $x_n$ are associated with different inequalities. For the $16$-QAM signaling, with $\beta_k = \beta_k^-$, the real and the imaginary parts of $x_n$ are associated with either one equality or an inequality depending on the data symbols; when $\beta_k > \beta_k^-$, at most two inequalities are introduced for either the real or the imaginary part. Therefore, the optimization problem of \eqref{Problem:general} can be generally reformulated as follows:
\begin{align}
{\bf \cal P}_1:\left\{ {\begin{array}{*{20}l}
   {\min \limits_{ \tilde{\mathbf{x}} \in \mathbb{R}^{2N\times1}} P\left({\bf \tilde{x}}\right) = \tilde{\mathbf{x}}^T\tilde{\mathbf{x}} }  \\
   {\text{s. t. \:\:} \mathbf{A}\tilde{\mathbf{x}} - \mathbf{c} \preceq \mathbf{0}, }\\
  {~\quad\quad \mathbf{B}\tilde{\mathbf{x}} - \mathbf{e} = \mathbf{0},}
  \end{array}} \right. \label{Problem:linear}
\end{align}
where for the 4-QAM system, we have $\mathbf{A} = \{a_{i,j}\}\in\mathbb{R}^{2K\times2N}$, $\mathbf{c}=\{c_{i}\}\in\mathbb{R}^{2K\times1}$, $\mathbf{B} = {\bf 0}$ and ${\bf e}=0$, while for the 16-QAM system, we have $\mathbf{A} = \{a_{i,j}\}\in\mathbb{R}^{4K\times2N}$, $\mathbf{c}=\{c_{i}\}\in\mathbb{R}^{4K\times1}$, $\mathbf{B}=\{b_{i,j}\}\in\mathbb{R}^{2K\times2N}$ and $\mathbf{e}=\{e_{i}\}\in\mathbb{R}^{2K\times1}$. The precise definitions of the matrices and the vectors
involved depend on constraint regions ${\cal C}(d_k)$, $\forall k$, {as defined in Section~\ref{subsubsec:4QAM} and Section~\ref{subsubsec:16QAM} for the 4-QAM and 16-QAM signalling, respectively}. Whether or not the inequality/equality constraints are active will depend on the transmitted data associated with the constellation. It is possible that some of the constraints may not be active, in which case the corresponding row entries of $\mathbf{A}$/$\mathbf{B}$ and $\mathbf{c}$/$\mathbf{e}$ are padded with zeros. In addition, matrices $\mathbf{A}$/$\mathbf{B}$ enjoy the same sparsity as the system matrix ${\bf H}$. In particular, when indexing matrices $\mathbf{A}$/$\mathbf{B}$, the index pair $\left(i,j\right)$ corresponds to a particular MS-subcarrier pair, and thus $a_{i,j}/b_{i,j} = 0$ whenever the particular subcarrier is not allocated to the MS.
\par {As an illustrative example, we consider a 16-QAM system with $N = 3$ subcarriers and $K = 2$ MSs. Each MS is allocated $L=2$ subcarriers: The first MS is allocated subcarriers $1$ and $2$, while the second MS is allocated subcarriers $2$ and $3$. The effective channel matrix ${\bf H}$ is assumed to be
\begin{align}
{\bf H} = \left[ {\begin{array}{*{20}c}
   {1+j} &  {-1+j} & {0}  \\
   {0} & {1+j} & {1-j}  \\
\end{array}} \right].
\end{align}
The data symbol intended for MS $1$ and $2$ is $d_1 = D_{16}$ and $d_2 = D_{11}$, respectively. Both MSs require the same SEP target $Pe$ and thus employ the same scaling factor $\beta = \beta_1 = \beta_2$. The noiseless received components at MSs are calculated as
\begin{align}
\left[ {\begin{array}{*{20}c}
   {{\bar y}_1 }  \\
   {{\bar y}_2}  \\
\end{array}} \right] = {\bf H}{\bf x} = \left[ {\begin{array}{*{20}c}
   {(1+j)x_1 + (-1+j)x_2}  \\
   {(1+j)x_2 + (1-j)x_3}  \\
\end{array}} \right].
\end{align}
Therefore, according to the constraint regions constructed (see Section~\ref{subsubsec:16QAM} and also Appendix~\ref{appendix:16QAM}), we have:}
\begin{align}
&\beta - \delta_0 \le {{\bar y}_1^{(r)}} \le \beta + \delta_0, ~~~~\beta - \delta_0 \le {{\bar y}_1^{(i)}} \le \beta + \delta_0, \\
&~~~~~~~~~~~~~~- {{\bar y}_2^{(r)}} \le -\left(2\beta - \sigma Q^{-1}\left(\sqrt{1- Pe}\right)\right), \nonumber\\
&~~~~~~~~~~~~~~- {{\bar y}_2^{(i)}} \le -\left(2\beta - \sigma Q^{-1}\left(\sqrt{1- Pe}\right)\right),
\end{align}
which, by simple algebra, can be further translated into a set of general linear constraints as in~\eqref{Problem:linear} with ${\bf B} = {\bf 0}$, ${\bf e} = {\bf 0}$,
{\footnotesize \begin{align}
&{\bf A} = \left[ {\begin{array}{*{20}c}
   {1} & {-1} & {-1} &{-1} &{0} &{0}  \\
   {-1}& {1} & {1} & {1} &{0} &{0} \\
   {1} & {1} & {1} & {-1} &{0} &{0} \\
   {-1} &{-1} &{-1} &{1} &{0} &{0} \\
   {0} &{0} &{-1} &{1} &{-1} &{-1} \\
   {0} &{0} &{0}&{0}&{0}&{0} \\
   {0}&{0}&{-1}&{-1}&{1}&{-1} \\
   {0} &{0} &{0}&{0}&{0}&{0} \\
\end{array}} \right] \nonumber\\
{\text{and}}~~&{\bf c} = \left[ {\begin{array}{*{20}c}
   {\beta + \delta_0}  \\
   {-(\beta - \delta_0)} \\
   {\beta + \delta_0} \\
   {-(\beta - \delta_0)}\\
   {-\left(2\beta - \sigma Q^{-1}\left(\sqrt{1- Pe}\right)\right)}\\
   {0}\\
   {-\left(2\beta - \sigma Q^{-1}\left(\sqrt{1- Pe}\right)\right)}\\
   {0}\\
\end{array}} \right].
\end{align}}
As we can observe, matrix ${\bf A}$ inherits sparsity from the effective channel matrix ${\bf H}$; transmission of $d_1 = D_{16}$ and $d_2 = D_{11}$ invokes four and two inequality constraints, respectively, and there is no strict equality constraint in this example.

\section{Precoding Algorithm Design with Parallel Computation Units} \label{sec:precoding:algorithm}
\par Problem ${\bf \cal P}_1$ is strictly convex and can be solved via a number of standard algorithms, such as the interior-point algorithm \cite{boyd2004convex}. Most of these algorithms are designed for centralized implementation and could be efficient enough for a small-scale problem. However, as the problem dimension increases, the computational complexity may be prohibitive. More importantly, the sparsity inherent to the problem may not be well exploited in standard solvers. Given these observations, we are interested in developing a precoding algorithm that leverages the sparsity to reduce complexity and is suitable for solving large-dimension problems using parallel computation units. These units may correspond to parallel processor cores (threads) at the BS computer \cite{zheng2014using} or parallel processors at the cloud to which the BS is connected \cite{CRAN2013}.

\subsection{Algorithm Design}
\par We now detail the precoding algorithm with a focus on the general problem as if all inequalities and equalities in \eqref{Problem:linear} were activated. The key technique is the dual decomposition approach, see, e.g., \cite{palomar2006tutorial}. Recalling the graph representation ${\cal G}$ introduced (see Fig.~\ref{fig:graph:representation}), we can map all PSNs and OSNs to parallel computation units.
\par We start with forming the Lagrangian function:
\begin{align}\label{Eq:lag}
{\cal L}\left(\mathbf{\tilde x},{\bm{\lambda}}, \bm{\nu}\right) = {\mathbf{ \tilde x}}^T\mathbf{ \tilde x} + {\bm{\lambda}}^T\left(\mathbf{A}\mathbf{\tilde x}-\mathbf{c}\right)+{\bm{\nu}}^T\left(\mathbf{B}\mathbf{ \tilde x}-\mathbf{e}\right),
\end{align}
where ${\bm{\lambda}} \in \mathbb{R}^{4K\times 1} \succeq 0 $ and ${\bm{\nu}} \in \mathbb{R}^{2K\times 1}$ are Lagrangian multipliers (dual variables), among which each pair of primal variables ${\mathbf{\tilde x}}_{[2n-1:2n]}$ is associated with PSN computation unit $n$ ($n=1,\dots,N$), while each tuple of dual variables $\{{\bm \lambda}_{[4k-3:4k]}, {\bm \nu}_{[2k-1:2k]}\}$ is associated with OSN computation unit $k$ ($k=1,\dots,K$). The dual problem is then defined as
\begin{equation}
\max_{\bm{\lambda},\bm{\nu}} g(\bm{\lambda},\bm{\nu}), ~~~~ \text{subject to\;} \bm{\lambda} \succeq 0, \label{Master_dual}
\end{equation}
with $g(\bm{\lambda},\bm{\nu})=\min_{\mathbf{ \tilde x}} \cal{L}\left(\mathbf{ \tilde x},\bm{\lambda}, \bm{\nu}\right)$ being the dual objective function.
\par Then one can solve the original problem by finding the optimal dual variables in an iterative manner. Specifically, at the $t$th iteration, for fixed dual variables $\bm{\lambda}^{(t)}$ and $\bm{\nu}^{(t)}$, to attain the minimization of Lagrangian, one sets the first-order derivative of the Lagrangian to zero, which leads to
\begin{equation}\label{Eq:sub}
\tilde{{\bf x}}^{*(t)} = -\frac{1}{2}\left(\mathbf{A}^T\bm{\lambda}^{(t)}+\mathbf{B}^T\bm{\nu}^{(t)}\right),
\end{equation}
or more explicitly,
\begin{align}\label{Eq:sub_separarte}
 &\tilde x_l^{*(t)}  =  - \frac{1}{2}\left( {\sum\limits_{i \in {\cal I}(\tilde x_l)} {a_{i,l} } } \lambda _i^{(t)}  + {\sum\limits_{i \in {\cal I}'(\tilde x_l)} {b_{i,l} } } \nu _i^{(t)}\right), \nonumber\\
 &\quad \quad \quad \quad \quad \quad \quad \quad \quad \quad \quad \quad l= 1, \ldots, 2N,
\end{align}
where ${\cal I}(\tilde x_l)$ and ${\cal I}'(\tilde x_l)$ denote the collection of indices of dual variables $\lambda$ and $\nu$ that have interactions with $\tilde x_l$, respectively, according to the graph ${\cal G}$. Note that PSN unit $n$ is in charge of computing the pair $\tilde{{\bf x}}^{*(t)}_{[2n-1:2n]}$, $n=1,\dots,N$. The corresponding dual function is given by:
\begin{align}
g(\bm{\lambda},\bm{\nu}) = {\bf{\tilde x}}^{*(t)^T } {\bf{\tilde x}}^{*(t)}  + {\bm \lambda} ^T ({\bf{A\tilde x}}^{*(t)}  - {\bf{c}}) + {\bm \nu} ^T ({\bf{B\tilde x}}^{*(t)}  - {\bf{e}}). \label{equ:dual:objective}
\end{align}
The dual variables are then updated by the OSN units in a parallel manner according to
\begin{align}\label{Eq:update_ag}
&\lambda_i^{(t+1)} = \left[\lambda_i^{(t)} + \frac{t-1}{t+2}\left(\lambda_i^{(t)} - \lambda_i^{(t-1)}\right) \right. \nonumber \\
& \quad\quad \left.+ \frac{1}{2\kappa}\left(\sum_{l \in {\cal I}(\lambda_i)}a_{i,l}{{\hat x}}^{(t)}_l - c_i\right)\right]^+, i=1,\ldots,4K, \nonumber \\
&\nu_j^{(t+1)} =\nu_j^{(t)} + \frac{t-1}{t+2}\left(\nu_j^{(t)} - \nu_j^{(t-1)}\right) \nonumber \\
&\quad\quad+ \frac{1}{2\kappa}\left(\sum_{l \in {\cal I}(\nu_j)}b_{j,l}\hat{{x}}^{(t)}_l - e_j\right), j=1,\ldots,2K,
\end{align}
where ${\cal I}(\lambda_i)$ and ${\cal I}(\nu_j)$ denote the collection of indices of primal variables that have interactions with $\lambda_i$ and $\nu_j$, respectively; notation $\left[.\right]^+$ denotes the projection onto the nonnegative orthant, $\kappa= \left(\|\bar{\mathbf{A}}\bar{\mathbf{A}}^T \|_1\|\bar{\mathbf{A}}\bar{\mathbf{A}}^T \|_{\infty}\right)^{1/2}$ with $\bar{\mathbf{A}} = [\mathbf{A}^T,\mathbf{B}^T]^T$ and
\begin{equation}
\hat{{x}}_l^{(t)} = \tilde{x}_l^{*(t)} + \frac{t-1}{t+2}\left(\tilde{x}_l^{*(t)} - \tilde{x}_l^{*(t-1)}\right), l= 1,\ldots,2N. \nonumber
\end{equation}
This dual-variable updating rule offers faster convergence speed than the conventional gradient updating~\cite{palomar2006tutorial} as shown in~\cite{giselsson2012accelerated}. The algorithm described is summarized in Table~\ref{table:algorithm}.

\begin{table}[t]
\begin{center} \caption{Precoding Algorithm with Parallel Computation Units} \label{table:algorithm}
\begin{tabular}{l}
\hline
1) {\bf Initialize}: dual variables $\bm{\lambda}^{(0)} > 0$, $\bm{\nu}^{(0)}$; \\
2) {\bf Repeat for $T$ iterations until convergence criterion is met}: \\
\quad 2.1)~{\bf For $n=1,\ldots,N$}: \\
\quad \quad~~PSN unit $n$ computes its primal variables ${\mathbf{\tilde x}}_{[2n-1:2n]}$ using (\ref{Eq:sub_separarte}), \\
\quad  \quad~~and broadcasts the updated values to its neighboring OSN units;\\
\quad 2.2)~{\bf For $k=1,\ldots,K$}: \\
\quad  \quad~~OSN unit $k$ computes its dual variables $\{{\bm \lambda}_{[4k-3:4k]}, {\bm \nu}_{[2k-1:2k]}\}$\\
\quad  \quad~~using (\ref{Eq:update_ag}) and broadcasts the updated values to its neighboring \\
\quad \quad ~~PSN units.\\\hline
\end{tabular}
\end{center}
\end{table}

\subsection{Complexity Analysis}

\par To quantify the complexity of the algorithm, we distinguish the communication overhead and computational complexity.

\par In the algorithm, to update its primal variables ${\mathbf{\tilde x}}_{[2n-1:2n]}$ via \eqref{Eq:sub_separarte}, PSN unit $n$ only needs to gather dual variables from its neighboring OSNs; therefore, the number of messages passed to PSN unit $n$ depends on the number of active dual variables and is at most ${4{\left| {{\cal I}(x_n)} \right|}}$ for the 16-QAM, and $2{\left| {{\cal I}(x_n)} \right|}$ for the 4-QAM. To update dual variables $\{{\bm \lambda}_{[4k-3:4k]}, {\bm \nu}_{[2k-1:2k]}\}$, OSN unit $k$ only needs to collect primal variables from its neighboring PSNs; therefore, the number of messages passed to OSN unit $k$ is at most $2{\left| {{\cal I}(y_k)} \right|}$ for both the 4/16-QAM. {Thus, with $T$ iterations, the total number of message-passings across computation units is ${\mathcal O}(4KLT)$ and ${\mathcal O}(6KLT)$ for the 4-QAM and 16-QAM system, respectively, where we have used the fact that $\sum\nolimits_{n = 1}^N{{\left| {{\cal I}(x_n )} \right|}}=\sum\nolimits_{k = 1}^K{{\left| {{\cal I}(y_k )} \right|}} = KL$ with $L$ being the number of non-zeroes in each signature.}

\begin{table*}[]
\centering
\caption{Computational complexity of the proposed algorithm}
\label{table:complexity}
\begin{tabular}{|l|c|c|c|c|}
\hline
\multicolumn{2}{|l|}{Schemes~\textbackslash~Operations} & ``$+$" / iteration & ``$\times$" / iteration & complexity ($T$ iterations) \\ \hline
\multirow{2}{*}{4-QAM}         & compute $(22):$        & $4KL-2N$        & $4KL$           & \multirow{2}{*}{${\mathcal O}(16KLT)$} \\ \cline{2-4}
                                        & compute $(24):$        & $12KL+10K$      & $8KL+4K$        &                           \\ \hline
\multirow{2}{*}{16-QAM}        & compute $(22):$        & $8KL-2N$        & $8KL$           & \multirow{2}{*}{${\mathcal O}(20KLT)$} \\ \cline{2-4}
                                        & compute $(24):$        & $12KL+10K$      & $8KL+4K$        & \\ \hline
\end{tabular}
%\vspace{-1.0em}
\end{table*}

\par {Table~\ref{table:complexity} summarizes the computational complexity for the algorithm proposed. To update each of its primal variables via \eqref{Eq:sub_separarte}, PSN unit $n$ needs at most $(4\left| {{\cal I}(x_n )} \right|-1)$ additions and $4\left| {{\cal I}(x_n )} \right|$ multiplications for the 16-QAM, and $(2\left| {{\cal I}(x_n )} \right|-1)$ additions and $2\left| {{\cal I}(x_n )} \right|$ multiplications for the 4-QAM. On the other hand, to update each of its dual variables via \eqref{Eq:update_ag}, OSN unit $k$ needs at most $(6\left| {{\cal I}(y_k )} \right|+5)$ additions and $(4\left| {{\cal I}(y_k )} \right|+2)$ multiplications for both the 4/16-QAM. Overall, the algorithm involves ${\mathcal O}(16KLT)$ additions and multiplications for the 4-QAM system, and involves ${\mathcal O}(20KLT)$ additions and multiplications for the 16-QAM system, where $T$ is the number of iterations. It is clear that the more sparse the signatures are, the less overall communication overhead and computational complexity are required to generate the precoded symbols in proposed algorithm.}

\par {For comparison, we note that the conventional ZF precoding of \eqref{equ:conventional:zf} has a computational complexity of ${\mathcal O}(\frac{8}{3}K^3 + 4NK^2)$ to compute precoding matrix ${\bf W} = {\bf H}^\dag({\bf H}{\bf H}^\dag)^{-1}$ and additional complexity of ${\mathcal O}(4KN)$ to generate each precoded symbol vector via ${\bf x} = {\bf W}{\bf d}$. Consider a transmission frame that consists of $T_{s}$ 4-QAM symbols intended for each MS and assume channel ${\bf H}$ remains unchanged during the frame. The ratio of the complexity of the proposed scheme to that of the ZF approach is thus quantified by $\rho = {\mathcal O}(16KLTT_s)/{\mathcal O}(\frac{8}{3}K^3 + 4NK^2 + 4KNT_s)$. It is clear that the smaller $L$, the smaller $\rho$ will be. In particular, $\rho \approx 4LT/K$ for a fully loaded system with $N=K$ and sufficiently large $T_s$. As an example with $K=32$, $L=4$ and $T=100$ iterations, ratio $\rho \approx 50$, which indicates the proposed scheme has approximately $50$ times the complexity of ZF precoding. Despite the increase in complexity, the proposed scheme is able to provide enormous transmit power reduction over ZF precoding and is more robust against imperfect channel state estimation, as will be shown later in Section~\ref{sec:results}.}

\section{Sparse MC-CDMA with Replica Constellations}\label{sec:replica:constellation}
\par In the previous sections, we have mainly focused on the system with standard QAM constellations for which convex constraints on the precoded vector are constructed according to the SEP targets. In this section, we assume the system adopts replica constellations, where each ${\cal D}_k$ is the periodic extension of a regular QAM constellation along the real and imaginary axes. We propose an optimized Tomlinson-Harashima Precoding (THP) under the SEP constraints by applying a similar approach we have used for the system with standard QAM constellations.

\subsection{THP-Basics}
\par We first briefly review some basic concepts related to THP, see, e.g., \cite{windpassinger2004precoding}.
\par
\begin{figure}[t]
\centering
\includegraphics[width=0.42\textwidth]{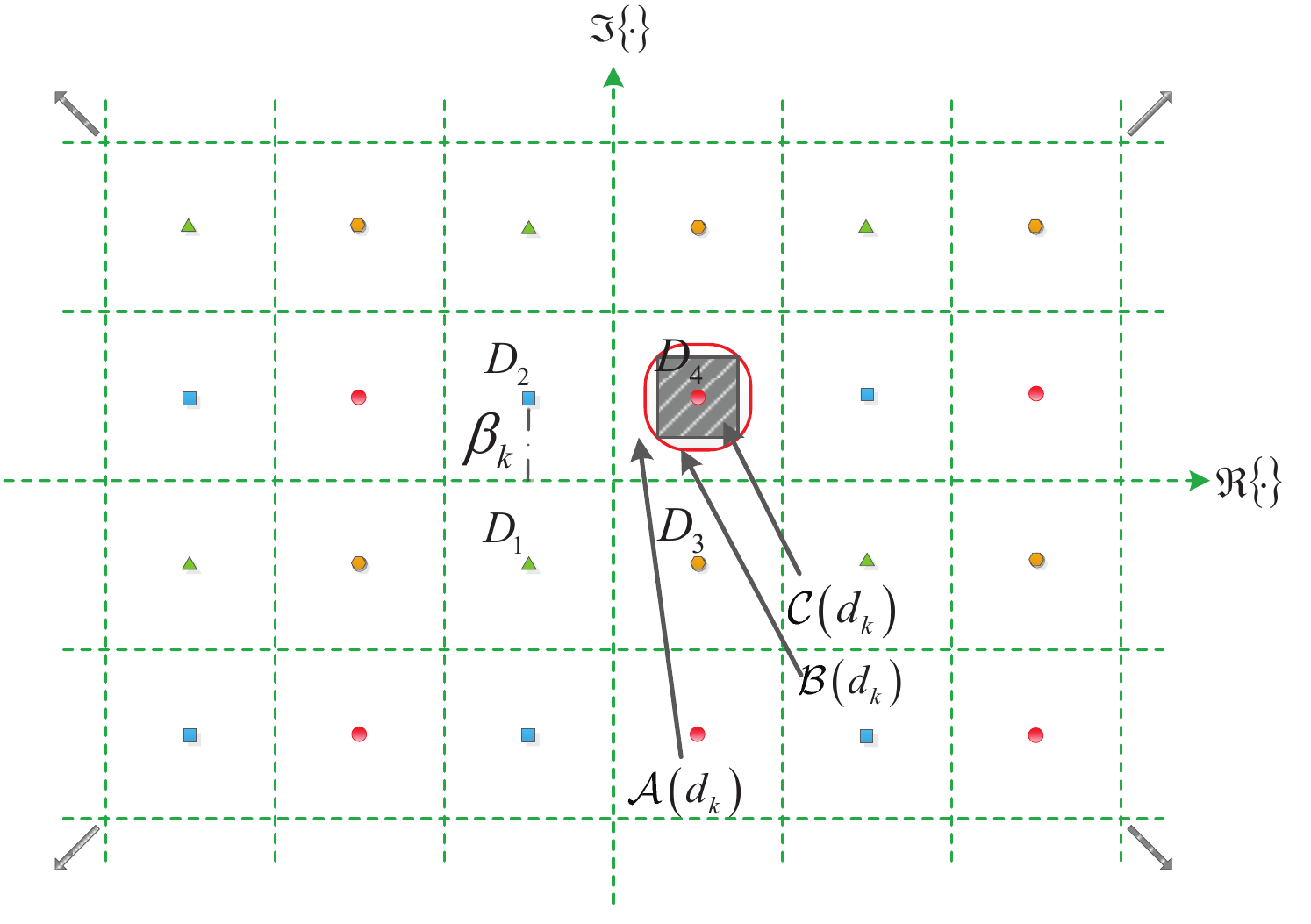}
\caption{Replica constellation built on 4-QAM: $\beta_k$ is a scaling factor; $\{D_1, D_2, D_3, D_4\}$ correspond to regular signal points before periodic extension; taking $d_k = D_4$ as an example, ${\cal A}(d_k)$ corresponds to the decision region, ${\cal B}(d_k)$ corresponds to the precise constraint region on noiseless output $\overline{y}_m$, while ${\cal C}(d_k)$ represents the constraint region with conservative approximation.}\label{fig:THP}
%\vspace{-1.5em}
\end{figure}

 In general, the replica constellation point ${d} \in {\cal D}_k$ can be represented as:
\begin{align}
{d  = D_l  + 2\beta _k \sqrt M \left( {a_{\text R}  + ja_{\text I} } \right)},
\end{align}
where $D_l$ corresponds to a regular point in the scaled $M$-QAM constellation under scaling $\beta_k$ ($l=1,\dots,M$), and $\left\{a_{\text R},a_{\text I}\right\}$ corresponds to an arbitrary integer pair, see Fig.~\ref{fig:THP} for a visual illustration when $M=4$. It is noted that decision regions associated with all replica points are identical closed squares with side length $2\beta_k$.
\par The THP is normally done in a successive manner in which interference created by previous users' transmissions is pre-cancelled to facilitate the transmission for the current user at each stage. The encoding is accommodated by the replica constellation and modulo-operation at the transmitter. Specifically, let the channel matrix be represented as $\mathbf{H}^\dag = \mathbf{F}\mathbf{R}$ as a result of QR factorization, where $\mathbf{F}$ is a unitary matrix and
$\mathbf{R}$ is an upper triangular matrix. Then $\mathbf{B} = \mathbf{H}\mathbf{F} = \mathbf{R}^\dag$ is a lower triangular matrix. The successive precoding operates as
\begin{equation} \label{Eq:THP}
\bar{x}_k =\frac{1}{\mathbf{B}_{k,k}} \left[ {d}_k - \sum\nolimits_{l=1}^{k-1}\mathbf{B}_{k,l}\bar{x}_l\right]_{p_k},
\end{equation}
where ${d}_k \in {\cal D}_k$ is the replica point carrying information for MS $k$ and $\left[ u \right]_{p_k}$ is the modulo operation operated on complex number $u$ with respect to basis ${p_k}$ and is defined as:
\begin{align}
\left[ u \right]_{p_k}= u - \left\lfloor {\frac{{\Re{\{u\}} + {p_k}/2}}{ {p_k}}} \right\rfloor {p_k} - j\left\lfloor {\frac{{\Im{\{u\}} + {p_k}/2}}{{p_k}}} \right\rfloor {p_k},
\end{align}
with ${p_k} = \sqrt{M}\beta_k$. The transmit signal is then formed by multiplying $\mathbf{F}$ with $\bar {\mathbf{x}}$, i.e., $\mathbf{x} = \mathbf{F}\bar{\mathbf{x}}$. In this way, at the receiver side, no MS experiences inter-user interference because of the pre-cancellation operations done at the BS. It is remarked that since THP is performed in a successive manner, different user orderings may lead to different performance. To find the optimal ordering, one needs to do an exhaustive search over all possible combinations, which is generally infeasible as $K$ goes large. In this work we simply adopt the suboptimal V-BLAST~(VB) ordering~\cite{golden1999detection}.

\subsection{Optimized THP}
\par Under the SEP constraints, we can formulate a precoding optimization problem similar to ${\cal P}_1$. The idea is that instead of choosing a minimum scaling factor $\beta_k^-$ for the constellation and performing a zero-forcing THP, one can relax the constraints on the noiseless output components and introduces room for optimizing the input signals as we scale up the constellation.
\par In particular, with the replica constellation scaled up, the resulting constraint regions become boxes centered at each replica point as constructed and approximated similarly for the inner points of the 16-QAM constellation, see Fig.~\ref{fig:THP}. The optimization problem is then formulated as:
\begin{align}
{\bf \cal P}_2:\left\{ {\begin{array}{*{20}l}
   {\min \limits_{ \tilde{\mathbf{x}} \in \mathbb{R}^{2N\times1}} P\left({\bf \tilde{x}}\right) = \tilde{\mathbf{x}}^T\tilde{\mathbf{x}} }  \\
   {\text{s. t. \:\:} \mathbf{A}\tilde{\mathbf{x}} - \mathbf{c} \preceq \mathbf{0},}\\
   \end{array}} \right.
\end{align}
where matrices $\mathbf{A} = \{a_{i,j}\}\in\mathbb{R}^{4K\times2N}$ and $\mathbf{c}=\{c_{i}\}\in\mathbb{R}^{4K\times1}$ are formed according to the constraints
\begin{align}
&{\cal C}(d_k) = \left\{ {\left( {\bar y_k^{(r)} ,\bar y_k^{(i)} } \right):\begin{array}{*{20}c}
   {d_k^{(r)}  - \delta _0  \le \bar y_k^{(r)}  \le d_k^{(r)}  + \delta _0 }  \\
   {d_k^{(i)}  - \delta _0  \le \bar y_k^{(i)}  \le d_k^{(i)}  + \delta _0 }  \\
\end{array}} \right\}, \nonumber\\
&\quad \quad \quad \quad \quad \quad \quad \quad \quad k = 1,\dots,K,
\end{align}
where parameter $\delta_0$ determines the size of the constraint box and is chosen to satisfy:
\begin{align}
{Q\left(\frac{\delta_0-\beta_k}{\sigma}\right) - Q\left(\frac{\delta_0 + \beta_k}{\sigma}\right)  = \sqrt{1-Pe_k}},
\end{align}
and the set of information-carrying replica points $\{{d}_k, k=1,\dots,K\}$ is determined from the ZF-THP encoding procedure. Therefore, the precoding algorithm proposed in Section~\ref{sec:precoding:algorithm} can be applied here to calculate the optimized THP precoded vector.

\section{Simulation Results}\label{sec:results}
\par We now present numerical results to demonstrate the effectiveness of the precoding schemes proposed for the sparse MC-CDMA system.
\subsection{Simulation Setup}
\par In the simulation, the noise variance $N_0$ is set to unity. The total number of subcarriers is fixed with $N=32$ and the number of MSs $K \le N$ is allowed to vary. Different MSs experience different frequency-selective fading channels. Specifically, the channel frequency response between the BS and MS $k$ is generated according to
\begin{align}
\tilde h_{k,n}  = \sum\limits_{q = 0}^{{\tilde Q} - 1} {g_{k,q} } e^{ - j\frac{{2\pi qn}}{N}} ,\:n = 1, \ldots ,N,
\end{align}
where ${\bf{g}}_k  = \left[ {g_{k,0} , \cdots, g_{k,{\tilde Q}- 1} } \right]^T$ represents the discrete-time channel response consisting of ${\tilde Q}$ taps; components $\{g_{k,q}\}$ are modeled as independent zero-mean Gaussian random variables, whose individual variance equals $\{\bar \lambda e^{-\frac{q}{4}}\}$ with normalization factor $\bar \lambda$ chosen such that ${\mathsf E}\left[ {\left\| {{\bf{g}}_k } \right\|^2 } \right] = 1$. In the simulation, ${\tilde Q}=8$ is adopted for the channel generation. {It is assumed that there is no inter-symbol-interference and inter-carrier-interference in the system.}
\par Unless stated otherwise, for any fixed system configuration, we simulate $1000$ transmission slots, under each of which random data and random channel are independently generated for each MS. In addition, we produce $10$ random regular signature matrix realizations as defined in Section~\ref{subsec:signature} and thus every $100$ transmission slots share the same signature matrix. The system transmitted power consumption presented shortly is averaged over all transmission slots. {For the precoding algorithm proposed, the calculation terminates at iteration $t$ if the normalized improvement of dual objective $\left| {g^{(t)}  - g^*} \right|/g^*  \le \delta =10^{-4}$, where $g^* \buildrel \Delta \over = \mathop {\max }\limits_{i \in \left\{ {1,...,t - 1} \right\}} g^{(i)}$}.

\subsection{System with Standard Constellation}
\par {Fix SEP target $Pe_k = 10^{-3}$ for all MSs and vary $K \in [24:32]$. In the sparse MC-CDMA, different levels of sparsity, e.g., $L=4$ and $L=8$, are considered. The case with $L=32$, referred to as the dense MC-CDMA, is also considered for the purpose of comparison}. {In addition, we have also compared with two conventional precoding schemes including ZF and the optimized RZF in form of:
\begin{align}
\mathbf{x}_{\rm ZF} &= \mathbf{H}^\dag\left(\mathbf{H}\mathbf{H}^\dag\right)^{-1} \mathbf{d},\\
\mathbf{x}_{\rm RZF} &= k_1\mathbf{H}^\dag\left(\mathbf{H}\mathbf{H}^\dag + k_2{\mathbf I}_K\right)^{-1}\mathbf{d},
\end{align}
where $\mathbf{d} =[d_1,\dots,d_K]^T$ with $d_k$ denotes the transmitted data symbol for MS $k$ and is drawn from a scaled version of standard constellation by $\beta^-_k$, ${\mathbf I}_K$ is a $K \times K$ identity matrix, and $k_1, k_2$ are two non-negative parameters to be optimized subject to the SEP constraints in~\eqref{Problem:linear}. {Note that the optimized RZF encompasses the conventional regularized ZF precoder~\cite{peel2005vector} with $k_1=1$ and also the minimum-mean-square-error (MMSE) precoder~\cite{bjornson2014optimal} with $k_1 =1$ and $k_2 = K\sigma^2/P.$}}

\begin{figure}[t]
\centering
\includegraphics[width = 0.4\textwidth]{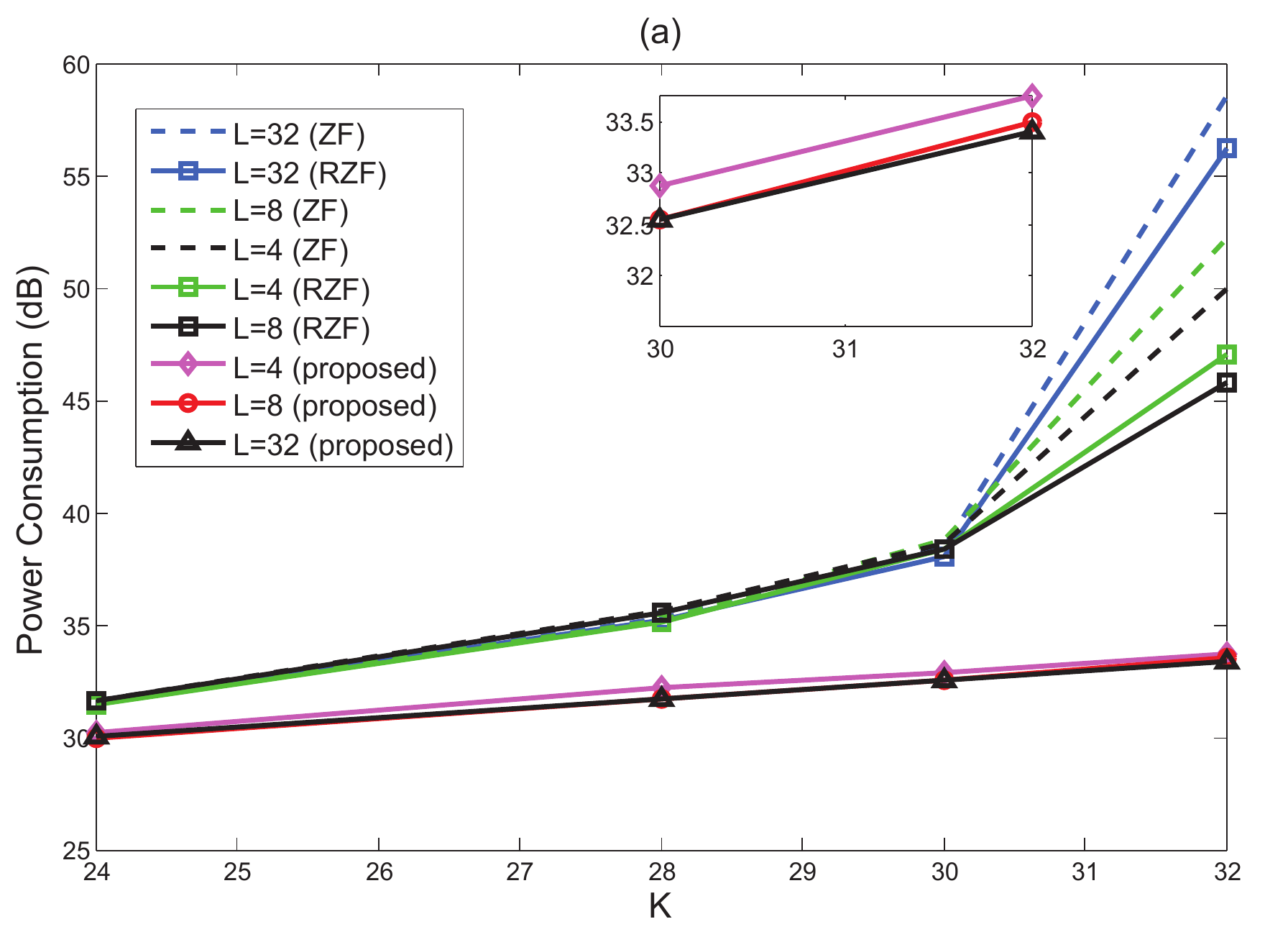}
\includegraphics[width = 0.4\textwidth]{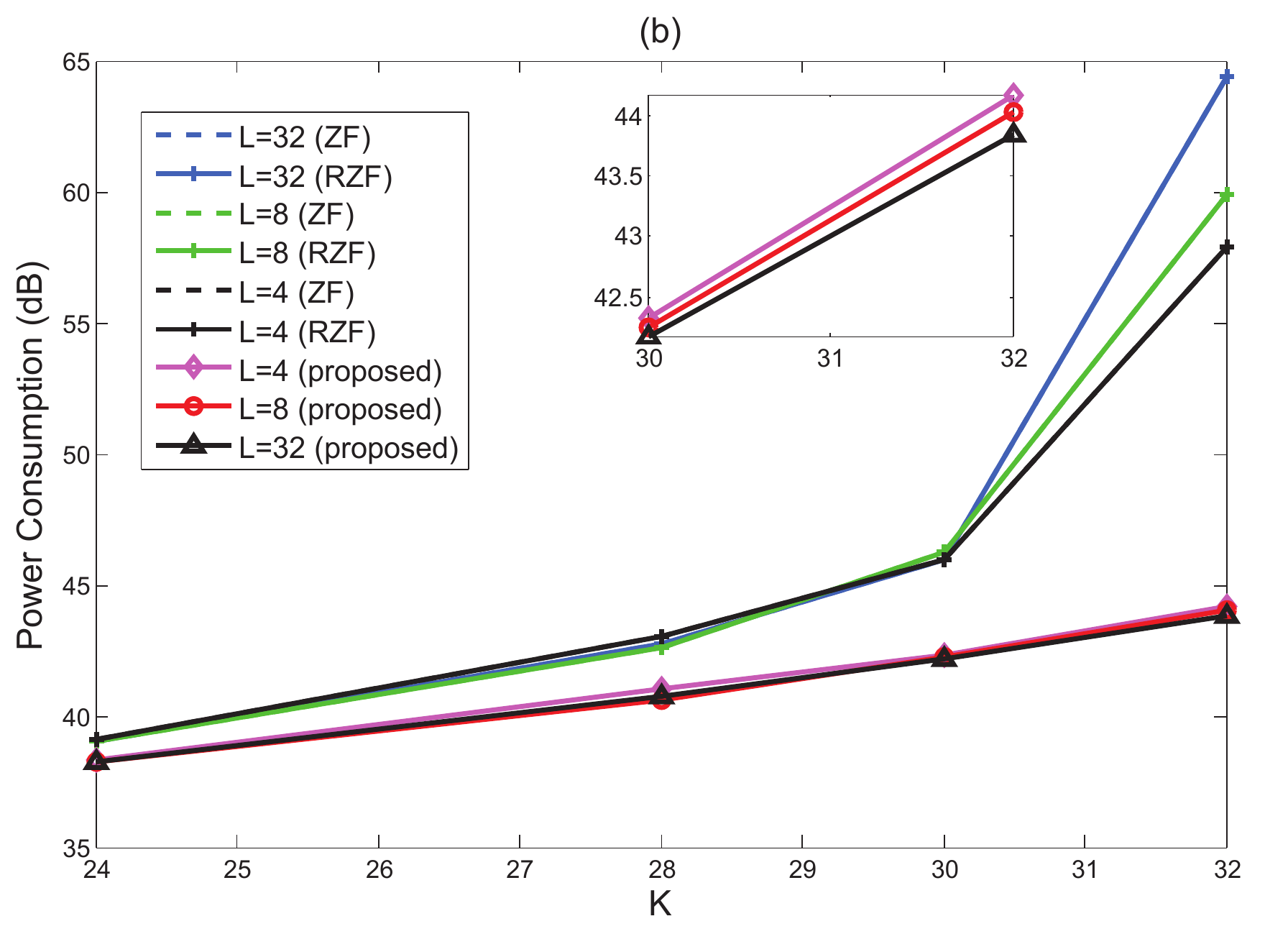}
\caption{Transmit power consumption at BS under different loads and different levels of sparsity ($N=32, Pe_k = 10^{-3}$): (a) standard 4-QAM; (b) standard 16-QAM.} \label{fig:4QAM:16QAM:power}
\end{figure}

\par Fig.~\ref{fig:4QAM:16QAM:power}-(a) and Fig.~\ref{fig:4QAM:16QAM:power}-(b) plot the transmit power consumption at the BS versus $K$ under different setups and precoding schemes for systems with 4-QAM and 16-QAM, respectively. Two important observations are made as follows.
\par First, power reduction from the proposed precoding over ZF {and RZF} is clearly evident for all load and sparsity combinations {considered}. In particular, for any fixed $L$, the reduction increases as $K$ grows. {For instance, when $K=N=32$ and $L=8$, for the 4-QAM, we have $18.8$ dB and $13.6$ dB reduction compared with ZF and RZF, respectively, while for the 16-QAM, we have $16$ dB reduction compared with both schemes, noting that the optimized RZF solutions are degraded and coincide with the ZF solutions in this case. As $K \to N$, the effective channel matrix ${\bf H}$ is increasingly likely to be poorly conditioned. Hence, the inefficiency of conventional schemes (in particular ZF) becomes pronounced. However, the precoding proposed is not sensitive to the conditional number of ${\bf H}$ and always attains the best performance}.
\par {Second, for the sparse MC-CDMA system, there is a trend that a denser signature (a larger value of $L$) leads to a smaller power consumption needed. For instance, the system with $L =4$ consumes slightly more power than the system with $L=8$ to achieve the same SEP target under both the 4-QAM and 16-QAM systems. However, to attain comparable power efficiency to the dense MC-CDMA, the signatures can still be relatively sparse ($L=8$ in our examples), yielding considerable reduction in precoding complexity. This observation also indicates that the sparse MC-CDMA system with proper choice of $L$ would attain almost the same link throughput as that of the dense MC-CDMA system under the same transmit power budget.}
\begin{figure}[t]
\centering
\includegraphics[width=0.4\textwidth]{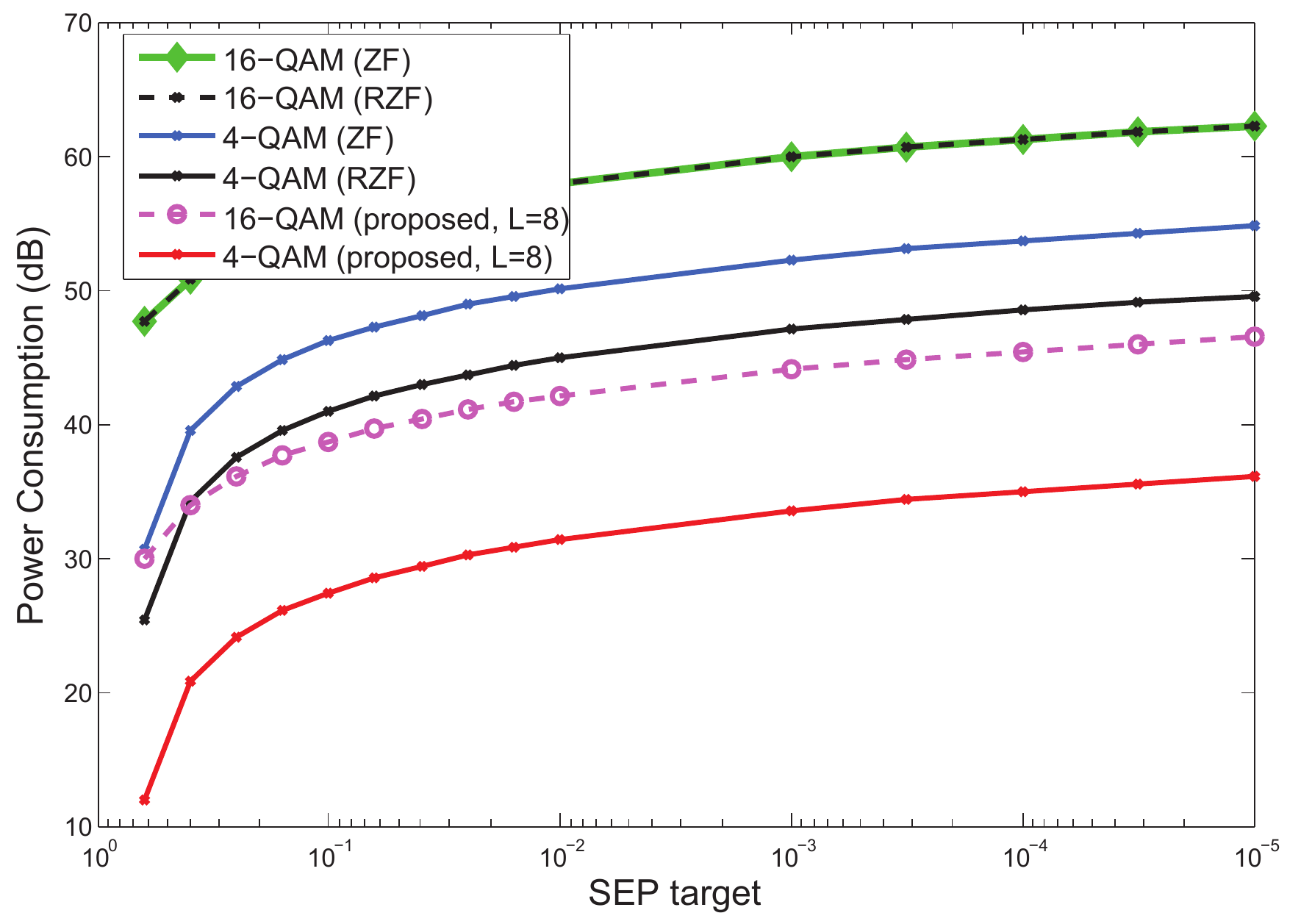}
\caption{{Power consumption versus different SEP targets ($K = N =32$, $L=8$).}}\label{fig:uncodedQAM:SEP}
\end{figure}
\par {Fig.~\ref{fig:uncodedQAM:SEP} plots the power consumption versus different SEPs with $K = N =32$ and $L=8$, which further confirms the superiority of the proposed scheme as compared to baselines ZF and RZF under different SEP targets for both 4-QAM and 16-QAM systems.}

\subsection{System with Replica Constellation}
\par We now consider the system with replica constellation, where system parameters $N=32$ and $L=8$.
 All MSs request the same minimum SEP target $Pe = 10^{-3}$. Under this SEP requirement, a uniform scaling factor across all MSs is chosen such that the power consumption is minimized for the proposed optimized THP. To perform the precoding optimization, we use an algorithm similar to that for systems with standard constellations. Therefore, signature sparsity is leveraged to reduce precoding complexity, as it was before.

\begin{figure}[t]
\centering
  \includegraphics[width=0.4\textwidth]{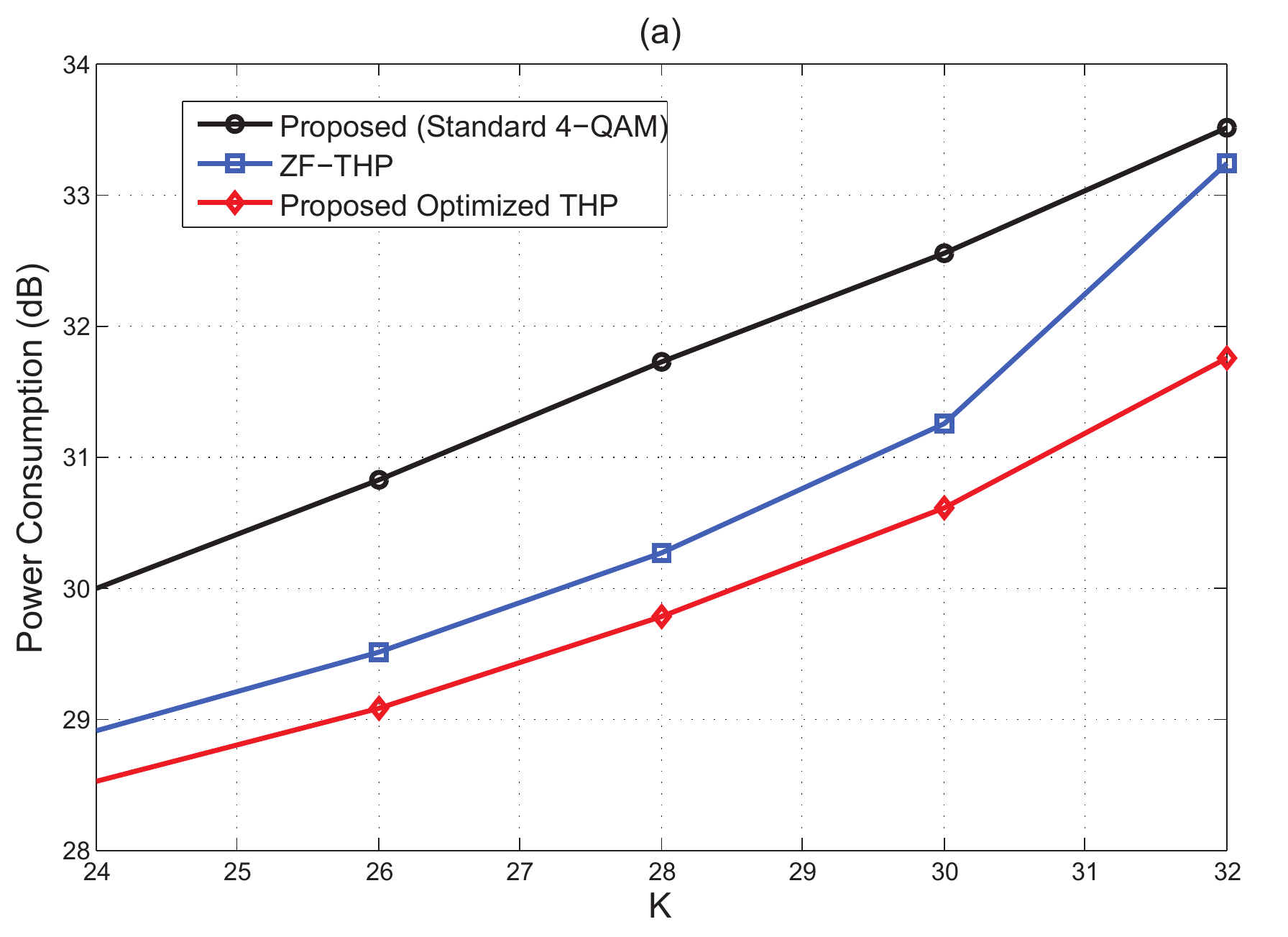}
  \includegraphics[width=0.4\textwidth]{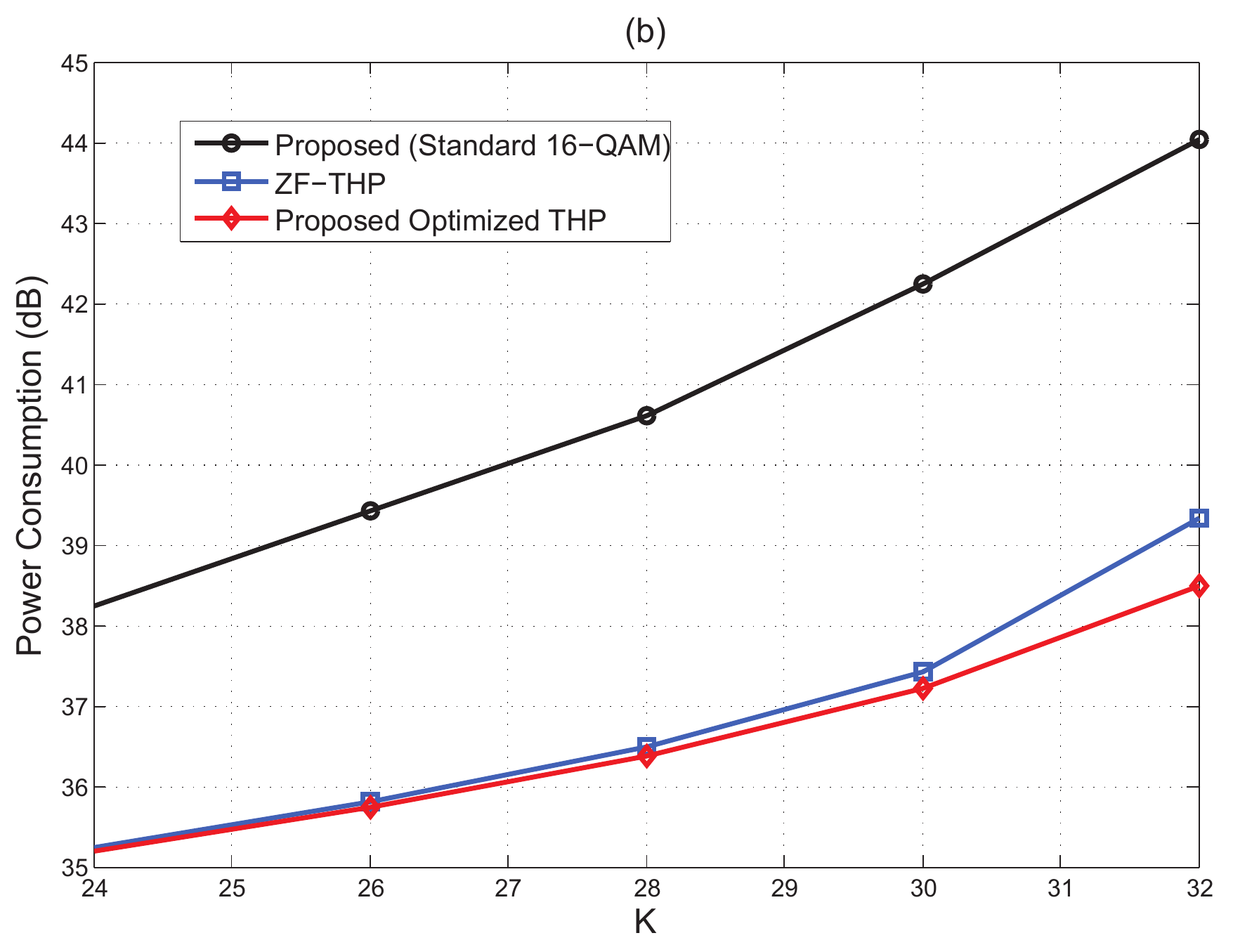}
 \caption{Power consumption for ZF-/optimized THP with replica points under different loads: (a) extended from standard 4-QAM; (b) extended from standard 16-QAM ($N=32, L=8, Pe_k = 10^{-3}$).} \label{fig:THP:4:16QAM}

\end{figure}

\par Fig.~\ref{fig:THP:4:16QAM}-(a) and Fig.~\ref{fig:THP:4:16QAM}-(b) plot the transmit power consumption versus $K$ under both the ZF- and optimized THP schemes for the system with 4-QAM and 16-QAM replica points, respectively. The performance of the proposed scheme under standard constellations is also included here for the purpose of comparison. It is seen that the optimized THP is able to provide significant power reduction over the proposed scheme under standard constellations. This further reduction, albeit appealing, does not come for free and has to be paid with more sophisticated encoding and decoding operations in THP schemes as described in Section~\ref{sec:replica:constellation}. It is also observed that the optimized THP generally outperforms ZF-THP in power efficiency and the exact gain depends on the system load. In particular, the former provides roughly $1.5$ dB power reduction over the latter for 4-QAM replica and roughly $0.85$ dB reduction for 16-QAM replica in a full-load system. The ZF-THP is already very power-efficient, yet the proposed THP is seen here to provide further reduction in transmit power.

\subsection{Bit Error Rate (BER) Results and Impact of Imperfect Channel Estimation}
\par
\begin{figure}[t]
 \centering
 \includegraphics[width=0.4\textwidth]{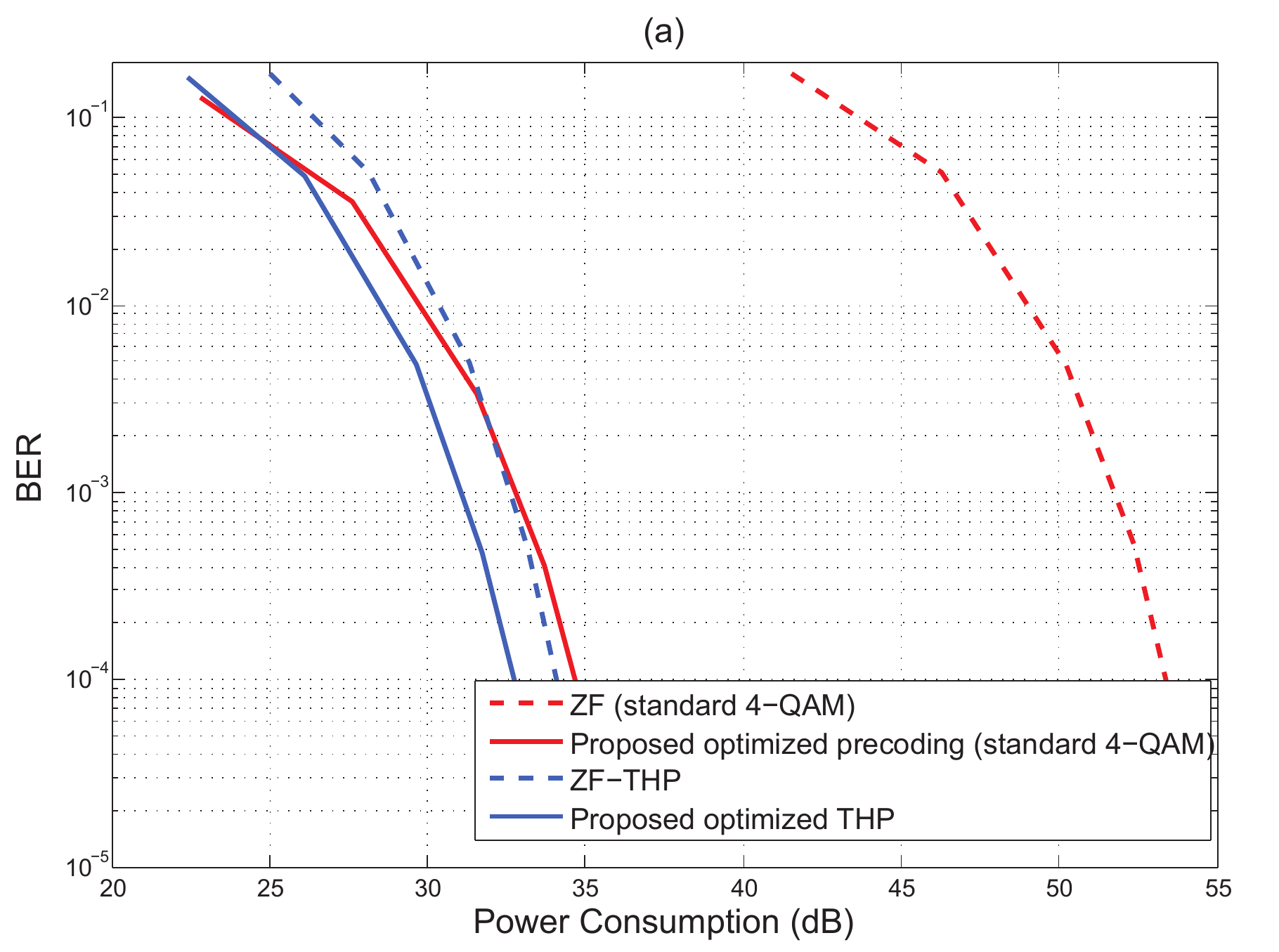}
  \includegraphics[width=0.4\textwidth]{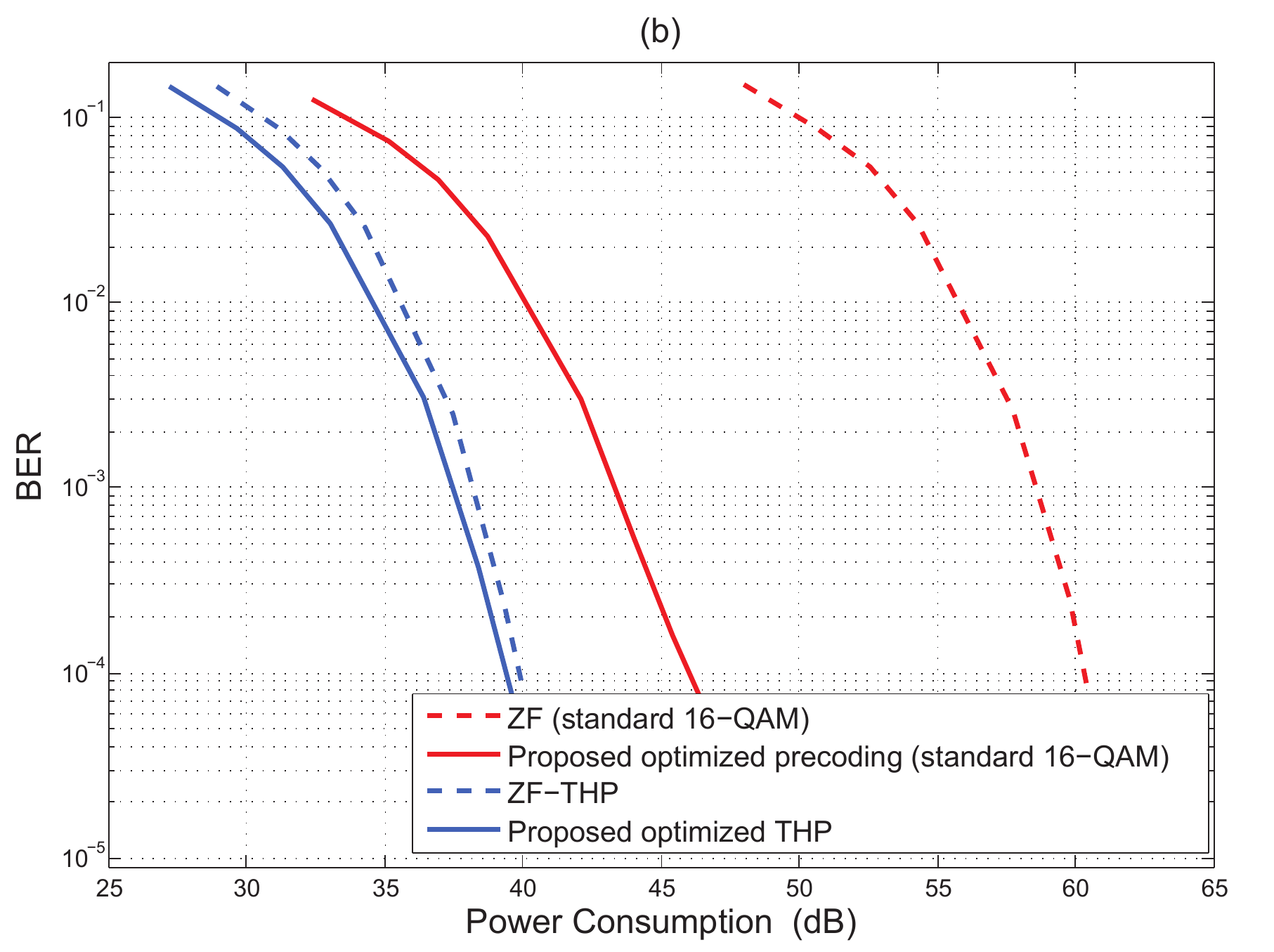}
 \caption{Uncoded BER versus power consumption ($K=N=32$, $L=8$): (a) standard and replica 4-QAM; (b) standard and replica 16-QAM.} \label{fig:BER:4:16QAM}
\end{figure}

 So far, we have demonstrated the power efficiency of the proposed precoding under different uncoded SEP targets. We now evaluate the impact of the proposed scheme on another practically important performance metric in terms of uncoded bit error rate (BER). The BER is calculated and averaged over $10^6$ realizations of transmissions. Fig.~\ref{fig:BER:4:16QAM} (a) and (b) depict the average BER (at a typical MS) as a function of power consumption for a system with standard/replica 4-QAM and 16-QAM, respectively. Consistent with the previous observations, the proposed optimized precoding significantly outperforms ZF precoding in terms of power efficiency to attain the same BER target under standard QAM constellations. The optimized THP is more power-efficient than the ZF-THP, and both of them generally outperform the optimized precoding under standard constellations but at the cost of increase complexity as explained before.

\par
\begin{figure}[t]
 \centering
 \includegraphics[width=0.4\textwidth]{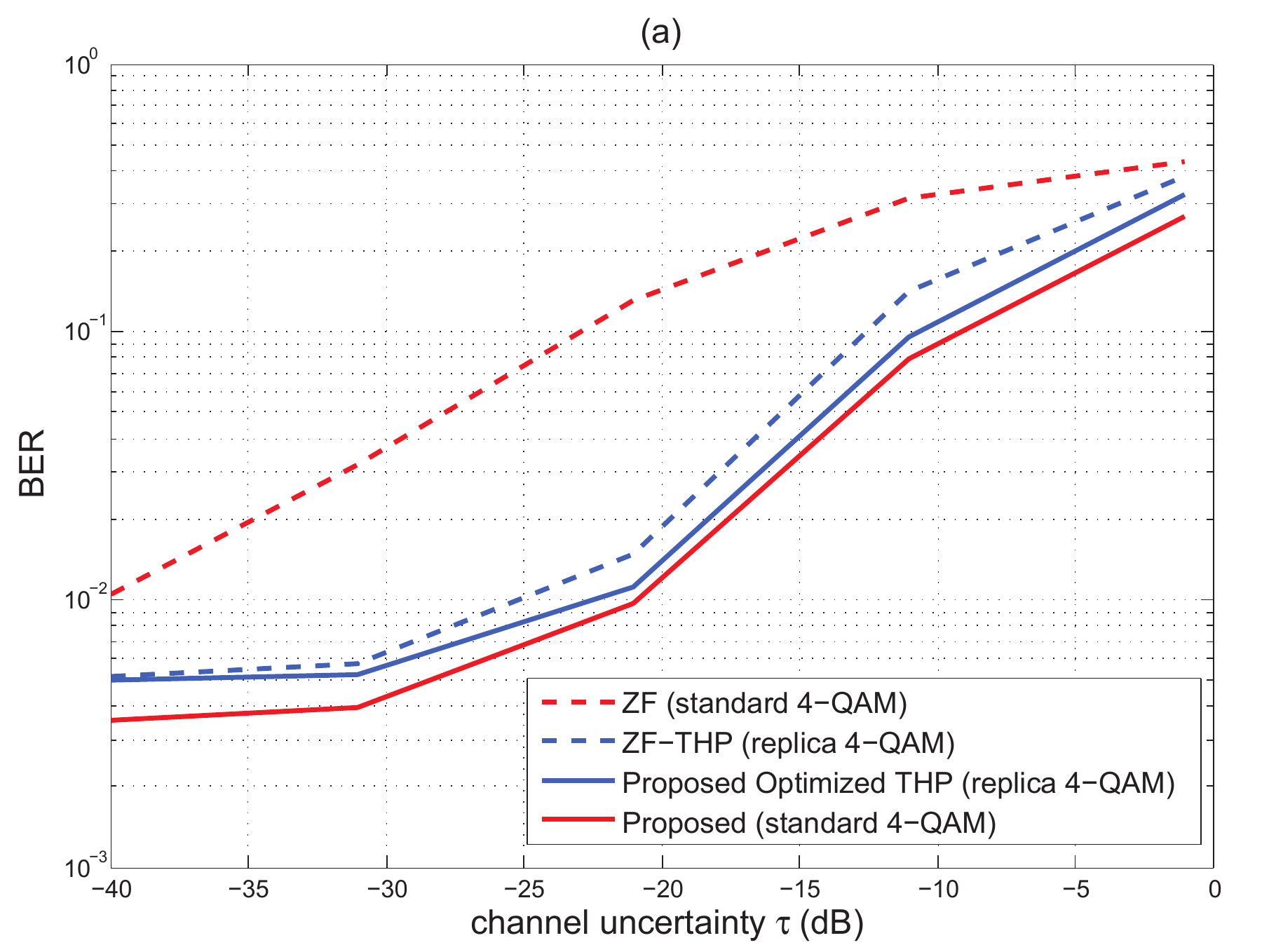}
 \includegraphics[width=0.4\textwidth]{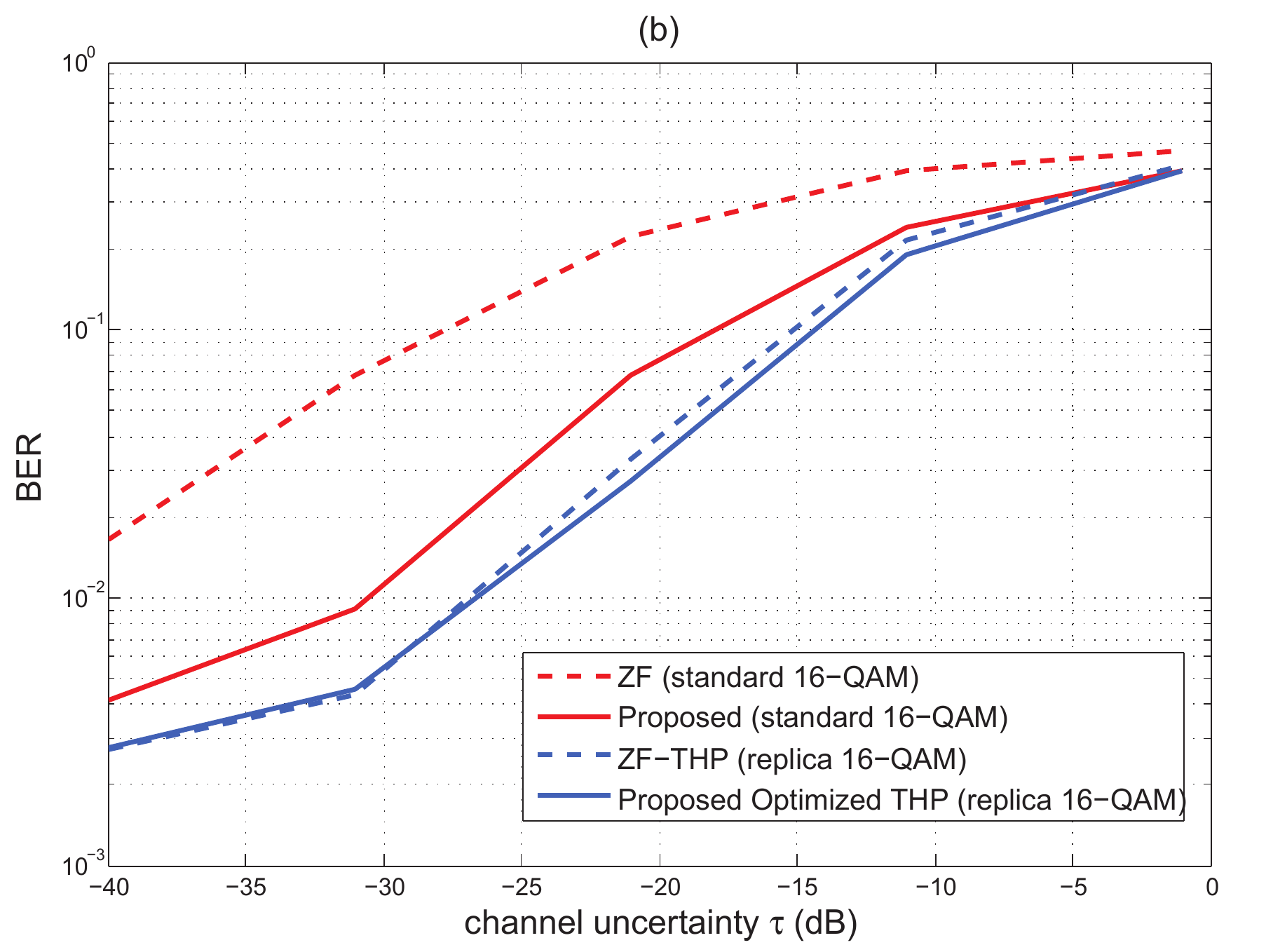}
\caption{{Uncoded BER versus the normalized channel uncertainty ($K=N=32$, $L=8$): (a) standard and replica 4-QAM; (b) standard and replica 16-QAM.}}\label{fig:BER:imperfectCSI}
\end{figure}

 Next, we evaluate the impact of channel estimation error (i.e., channel uncertainty) on the performance of the schemes considered. Let ${\bf {\hat H}}$ denote the estimated sparse channel matrix. Each nonzero entry ${\hat h}_{k,n}$ of ${\bf {\hat H}}$ is a noisy version of the perfect ${h}_{k,n}$ of ${\bf {H}}$. To model the uncertainty, we assume ${\hat h}_{k,n}$ is generated according to: ${\hat h}_{k,n} = {h}_{k,n} + z_{k,n}$, where $z_{k,n} \sim {\cal CN}(0,\sigma_{e}^2)$ represents the complex Gaussian estimation error with variance $\sigma_{e}^2$ and $\{z_{k,n}, \forall k, \forall n\}$ are independently and identically distributed. The average normalized channel uncertainty is then defined as $\tau = {\mathsf E}[10\log_{10}(\|{\bf \hat H} - {\bf H}\|_2^2/\|{\bf H}\|_2^2)]$ in dB. For all schemes evaluated, the SEP target $Pe_k$ is set to $10^{-2}$ so that the corresponding BER is on the order of ${10^{-3}}$, if perfect channel state information is available. Fig.~\ref{fig:BER:imperfectCSI} (a) and (b) depict the real BER (at a typical MS) versus different levels of channel uncertainty for a sparse MC-CDMA system with standard/replica 4-QAM and 16-QAM signaling, respectively. It can be seen that as the channel uncertainty increases, the real BER of both the ZF approach and the proposed precoding scheme degrade. However, the proposed scheme always outperforms its ZF counterpart and exhibits much better robustness against imperfect channel estimation, particularly for a system with standard constellations. The proposed scheme is thus not only more power-efficient but also more robust against channel uncertainty.

\section{Conclusions} \label{sec:conclusions}

\par In this work, we have introduced a sparse MC-CDMA downlink model and proposed a power-efficient precoding method under minimum symbol error probability (SEP) requirements at the MSs. It has been shown that when the system load is high, the proposed precoder significantly reduces the transmission power over (regularized) zero-forcing based precoders under the same SEP targets. It has also been shown that using the proposed precoder, the sparse MC-CDMA system {with a proper choice of sparsity level} attains almost the same power efficiency {and link throughput} as that of the dense MC-CDMA, but with lower complexity. These features, along with the fact that channel measurements may be simplified with sparse signatures, add to the practical appeal of the sparse MC-CDMA and make it a valuable candidate for future-generation wireless communication systems.

\appendices
\section{Critical Points on the Boundary of ${\cal B}(d_k)$ for the 16-QAM Signaling}\label{appendix:16QAM}
\par For the 16-QAM signaling, the SEP target \eqref{Eq:SEP} indicates that
\begin{eqnarray} \label{Cons:16:QAM}
&\underbrace{\frac{1}{\sqrt{2\pi}\sigma}\int_{d_k^{(r)}-\overline{y}_k^{(r)}-\rho^{(r)}_{-}}^{d_k^{(r)}-\overline{y}_k^{(r)}+\rho^{(r)}_{+}}e^{-\frac{\left( z^{(r)}_k \right)^2}{2\sigma^2}} dz^{(r)}_k }_{\mathcal{O}^{(r)}}  \nonumber\\
&\times \underbrace{\frac{1}{\sqrt{2\pi}\sigma}\int_{d_k^{(i)}-\overline{y}_k^{(i)}-\rho^{(i)}_{-}}^{d_k^{(i)}-\overline{y}_k^{(i)}+\rho^{(i)}_{+}}e^{-\frac{\left( z^{(i)}_k \right)^2}{2\sigma^2}} dz^{(i)}_k}_{\mathcal{O}^{(i)}} \geq 1-Pe_k,
\end{eqnarray}
where $\rho^{(r)}_{+}/ \rho^{(i)}_{+}$ and $\rho^{(r)}_{-}/\rho^{(i)}_{-}$ relate to the decision regions and parameterize the upper/lower bounds for the integrals. Table~\ref{Table:limits} specifies these parameters for different constellation points.

\begin{table}[t]
\caption{Integral Parameters for the 16-QAM Signaling}
\begin{center}
{\begin{tabular}{|c|c|c|c|}
\hline
\backslashbox{$\rho^{(r)}_{-}/ \rho^{(r)}_{+}$}{$\rho^{(i)}_{-}/ \rho^{(i)}_{+}$} & $+\infty/\beta_k$ &$\beta_k/\beta_k$ & $\beta_k/+\infty$  \\
\hline
$+\infty/\beta_k$ & $D_1$ & $D_{\{2,4\}}$ &$D_3$\\
\hline
$\beta_k/\beta_k$ & $D_{\{5,13\}}$ & $D_{\{6,8,14,16\}}$ & $D_{\{7,15\}}$\\
\hline
$\beta_k/+\infty$ & $D_9$ &$D_{\{10,12\}}$ &$D_{11}$ \\
\hline
\end{tabular}}
\label{Table:limits}
\end{center}
%\vspace{-2.8em}
\end{table}

\par Given symbol $d_k \in {\cal D}_k$ and a target $Pe_k$, one can determine the precise constraint region ${\cal B}(d_k)$ on noiseless output $\overline{y}_k $ at MS $k$ from inequality~\eqref{Cons:16:QAM}. In particular, the boundary of the region is determined by the equality ${\mathcal{O}^{(r)}}{\mathcal{O}^{(i)}} = 1 - Pe_k$ in~\eqref{Cons:16:QAM}. In the following, we explain how to determine a set of critical points on the boundary of the corresponding regions for three representative constellation points $\{D_{11}, D_{12}, D_{16}\}$.
\par For the corner constellation point $D_{11}$, the critical boundary points can be determined according to three different combinations of $\left(\mathcal{O}^{(r)}, \mathcal{O}^{(i)}\right)$:
\begin{enumerate}[(i)]
\item{$\left(1-Pe_k, 1\right)$: $\overline{y}^{(r)}_k = 2\beta_k- {\sigma}{Q}^{-1}(1-Pe_k)$, $\overline{y}^{(i)}_k = +\infty$;}
\item{$\left(1, 1-Pe_k\right)$: $\overline{y}^{(r)}_k = +\infty$, $\overline{y}^{(i)}_k = 2\beta_k - {\sigma}{Q}^{-1}(1-Pe_k)$;}
\item{$\left(\sqrt{1-Pe_k},\sqrt{1-Pe_k}\right)$} : \\$~~~~~~~~~~~~~~~~~~~~\overline{y}^{(r)}_k = \overline{y}^{(i)}_k = 2\beta_k - {\sigma}{Q}^{-1}\left(\sqrt{1-Pe_k}\right)$,
\end{enumerate}
where $Q^{-1}(.)$ denotes the inverse of the standard $Q$-function.
\par For the center constellation point $D_{16}$, the critical boundary points can be found by examining the following combinations of $\left(\mathcal{O}^{(r)}, \mathcal{O}^{(i)}\right)$:
\begin{enumerate}[(i)]
\item{$\left(\frac{1-Pe_k}{\alpha},\alpha\right)$: $\overline{y}^{(r)}_k = \beta_k \pm \delta_1$, $\overline{y}^{(i)}_k = \beta_k$;}
\item{$\left(\alpha, \frac{1-Pe_k}{\alpha}\right)$: $\overline{y}^{(r)}_k = \beta_k$, $\overline{y}^{(i)}_k = \beta_k \pm \delta_1$;}
\item{{$\left(\sqrt{1-Pe_k},\sqrt{1-Pe_k}\right)$}: $\overline{y}^{(r)}_k = \beta_k \pm \delta_0$, $\overline{y}^{(i)}_k = \beta_k \pm \delta_0$,}
\end{enumerate}
where $\alpha= \frac{1}{\sqrt{2\pi}\sigma}\int_{-\beta_k}^{\beta_k}e^{-\frac{v^2}{2\sigma^2}}dv \geq \sqrt{1-Pe_k}$, parameters $\delta_0$ and  $\delta_1$ are chosen to satisfy:
\begin{align}\label{Eq:width}
\left\{ {\begin{array}{*{20}l}
   {Q(\frac{\delta_0-\beta_k}{\sigma}) - Q(\frac{\delta_0 + \beta_k}{\sigma})  = \sqrt{1-Pe_k}, }  \\
   {Q(\frac{\delta_1-\beta_k}{\sigma}) - Q(\frac{\delta_1 + \beta_k}{\sigma}) = \frac{1-Pe_k}{\alpha}.}\\
\end{array}} \right.
\end{align}
\par For the side constellation point $D_{12}$, the critical boundary points can be found similarly by examining the following combinations of $\left(\mathcal{O}^{(r)},\mathcal{O}^{(i)}\right)$:
\begin{enumerate}[(i)]
\item{$\left(\alpha, \frac{1-Pe_k}{\alpha}\right)$: $\overline{y}^{(r)}_k = 2\beta_k - {\sigma}{Q}^{-1}\left( \frac{1-Pe_k}{\alpha}\right)$, $\overline{y}^{(i)}_k = \beta_k$;}
\item $\left(\sqrt{1-Pe_k},\sqrt{1-Pe_k}\right)$: \\
$~~~~~~\overline{y}^{(r)}_k = 2\beta_k - {\sigma}{Q}^{-1}\left(\sqrt{1-Pe_k}\right)$, $\overline{y}^{(i)}_k = \beta_k \pm \delta_0$;
\item{{$\left(1, 1-Pe_k\right)$}: $\overline{y}^{(r)}_k = +\infty$, $\overline{y}^{(i)}_k = \beta_k \pm \delta_2$,}
\end{enumerate}
where parameter $\delta_2$ is chosen to satisfy:
\begin{align}
Q\left(\frac{\delta_2-\beta_k}{\sigma}\right) - Q\left(\frac{\delta_2 + \beta_k}{\sigma}\right) = 1-Pe_k.
\end{align}

\bibliographystyle{IEEETran}
% Generated by IEEEtran.bst, version: 1.12 (2007/01/11)

\end{document}